\documentclass[
amsmath,amssymb,
aps,
prb,
twocolumn,
]{revtex4}

\usepackage{graphicx}
\usepackage{dcolumn}
\usepackage{bm}
\usepackage{hyperref}
\usepackage{multirow}
\usepackage{array}
\usepackage{booktabs}
\usepackage{ctable}
\usepackage{upgreek}
\usepackage{epsfig,psfrag,subfigure,amsopn}
\usepackage{mathrsfs}
\usepackage{amssymb}
\usepackage{amsbsy}
\usepackage{color}
\usepackage{cancel}
\usepackage{pifont}
\usepackage{marginnote}
\usepackage{float}
\newcommand{\bcen}{\begin{center}}
\newcommand{\ecen}{\end{center}}
\newcommand{\btab}{\begin{tabular}}
\newcommand{\etab}{\end{tabular}}
\newcommand{\bdes}{\begin{description}}
\newcommand{\edes}{\end{description}}

\newcommand{\beq}{\begin{equation}}
\newcommand{\eeq}{\end{equation}}
\newcommand{\bea}{\begin{eqnarray}}
\newcommand{\eea}{\end{eqnarray}}
\newcommand{\non}{\nonumber}

\newcommand{\bary}{\begin{array}}
\newcommand{\eary}{\end{array}}
\newcommand{\benum}{\begin{enumerate}}
\newcommand{\eenum}{\end{enumerate}}
\newcommand{\bitem}{\begin{itemize}}
\newcommand{\eitem}{\end{itemize}}

\newcommand{\al}{\alpha}
\newcommand{\de}{\delta}
\newcommand{\ep}{\epsilon}
\newcommand{\ga}{\gamma}
\newcommand{\lam}{\lambda}
\newcommand{\om}{\omega}
\newcommand{\si}{\sigma}
\newcommand{\dg}{\dagger}

\newcommand{\bi}{\bibitem}

\newcommand{\Fig}[1]{Fig.~\ref{#1}}

\makeatletter

\newcommand{\Rmnum}[1]{\expandafter\@slowromancap\romannumeral #1@}
\makeatother

\newcommand{\mylabel}[1]{\label{#1}} 

\begin{document}
\relax



\title{Effects of local periodic driving on transport and generation of 
bound states}

\author{Adhip Agarwala$^1$ and Diptiman Sen$^2$}
\affiliation{\small{$^1$Department of Physics, Indian Institute of Science,
Bengaluru 560012, India \\
$^2$Centre for High Energy Physics, Indian Institute of Science, Bengaluru
560012, India}}



\date{\today}

\begin{abstract}
We periodically kick a local region in a one-dimensional lattice and 
demonstrate, by studying wave packet dynamics, that the strength and the time 
period of the kicking can be used as tuning parameters to control the 
transmission probability across 
the region. Interestingly, we can tune the transmission to zero 
which is otherwise impossible to do in a time-independent system. We adapt the 
non-equilibrium Green's function method to take into account the effects of 
periodic driving; the results obtained by this method agree with those found 
by wave packet dynamics if the time period is small. We discover that Floquet 
bound states can exist in certain ranges of parameters; when the driving 
frequency is decreased, these states get delocalized and turn into resonances 
by mixing with the Floquet bulk states. We extend these results to 
incorporate the effects of local interactions at the driven site, and we find 
some interesting features in the transmission and the bound states.
\end{abstract}

\maketitle

\section{Introduction}
\label{sec_intro}

Periodically driven quantum systems have attracted an immense amount
of interest for many years. A large variety of interesting phenomena 
resulting from periodic driving have been discovered including the coherent 
destruction of tunneling~\cite{grossmann91,kayanuma94}, the generation of 
defects~\cite{mukherjee}, dynamical freezing~\cite{das10}, dynamical 
saturation~\cite{russomanno12} and localization~\cite{alessio,nag,agarwala16},
dynamical fidelity~\cite{sharma14}, edge singularity in the probability
distribution of work~\cite{russomanno15} and thermalization~\cite{lazarides14a}
(for a review see Ref.~\onlinecite{dutta15}). There have also been studies of 
periodic driving of graphene by the application of electromagnetic 
radiation~\cite{gu11,kitagawa11,morell12,sentef15}, Floquet topological phases 
of matter and the generation of topologically protected states at the
boundaries~\cite{kitagawa10,lindner11,jiang11,trif12,gomez12,dora,liu13,tong13,
rudner13,katan13,lindner13,kundu13,basti13,schmidt13,reynoso13,wu13,manisha13,
perez,reichl14,manisha14,claassen16,sid17,hubener17}. Some of these aspects 
have been experimentally studied~\cite{kitagawa12,rechtsman,tarruell12,jotzu14}.

In addition, there have been several studies of the effects of interactions 
between electrons in periodically driven systems~\cite{eckardt05,rapp12,
zheng14,greschner14,lazarides14bc,rigol14,ponte,eckardt15,bukov16,lazarides14d,
keyser16,else,itin15}. The effects of interactions in Floquet topological 
insulators have been studied in Ref.~\onlinecite{mikami16}. It is known that 
interactions can lead to a variety of topological phases (some of which
have elementary excitations with fractional charges) in driven Rashba 
nanowires~\cite{klino16,klino17}, and to a chaotic and topologically trivial 
phase in the periodically driven Kitaev model~\cite{su16}. The effects of 
periodic driving on the stability of a bosonic fractional Chern insulator has 
been investigated~\cite{raciunas16}. Interestingly some of these systems have 
been realized experimentally demonstrating correlated hopping in the Bose 
Hubbard model~\cite{meinert2016} and many-body localization~\cite{bordia}, 
and realizing bound states for two particles in driven photonic 
systems~\cite{mukherjee16}.

Periodic driving can lead to an interesting phenomenon called dynamical 
localization. Here the particles become perfectly localized in space due to 
periodic driving of some parameter in the Hamiltonian. Systems exhibiting 
dynamical localization include driven two-level systems~\cite{grossmann91}, 
classical and quantum kicked 
rotors~\cite{chirikov81,fishman82,ammann98,tian11,nieuwenburg12}, the Kapitza
pendulum~\cite{kapitza51,broer04}, and bosons in an optical 
lattice~\cite{horstmann07}. It has been shown that remnants of dynamical
localization may survive even in the presence of strong disorder~\cite{roy15}.

In an earlier paper, it was shown that a combination of interactions and 
periodic $\de$-function kicks with a particular strength on all the sites 
on one sublattice of a one-dimensional system can lead to the formation of 
multi-particle bound states in three different models~\cite{agarwala17}. These
bound states are labeled by a momentum which is a good quantum number since
the system is translation invariant. This naturally leads us to ask if 
periodic kicks applied to only one site in a system can also lead to the 
formation of a bound state which is localized near that particular site.
Further, it would be interesting to the effect of such a localized periodic
kicking on the transmission across the site; a similar analysis for 
localized harmonic driving has been carried out in 
Refs.~\onlinecite{reyes,thuberg}. One can also study what happens if there 
is both a time-independent on-site potential (which can produce a 
bound state and affect the transmission on its own) and periodic kicking 
at the same site. Finally, one can study what the combined effect is of an
interaction (between, say, a spin-up and a spin-down electron) and periodic 
kicking at the same site. We will study all these problems in this paper.

In one dimension it is known that periodic driving in a local region 
can lead to charge pumping; see Refs.~\onlinecite{agarwal07,soori10} and
references therein. This is a phenomenon in which a net charge moves 
in each time period between two leads which are connected to the left and 
right sides of the region which is subjected to the driving. Charge pumping
can happen even when no voltage bias is applied between the leads; however, 
this requires a breaking of left-right symmetry which can only occur if the 
periodic driving is applied to more than one site. In this paper, we will 
study the effect of driving at only site; this cannot produce charge pumping.

The plan of this paper is as follows. In Sec.~\ref{sec_ham}, we will introduce
the basic model. We will consider a tight-binding model with spinless electrons 
in one dimension where periodic $\de$-function kicks are applied to the 
potential at one particular site. The strength and time period of the kicks
will be denoted by $\al$ and $T$ respectively. In Sec.~\ref{sec_trans}, 
we will discuss wave packet dynamics and how this can be used to compute the 
reflection and transmission probabilities across the site which is subjected 
to the periodic kicks. In Sec.~\ref{sec_dynloc}, we will discuss why there is 
perfect reflection from the kicked site for a particular value of $\al$ and 
how this is related to dynamical localization. In Sec.~\ref{sec_negf}, we 
will show how an effective Hamiltonian can be defined and will use this 
to calculate the zero temperature differential conductance (which is 
related to the transmission probability) using the non-equilibrium Green's 
function method~\cite{datta}. We will see that this matches the result 
obtained by the wave packet dynamics if $T$ is less than some value.
In Sec.~\ref{sec_bound}, we will discuss how the periodic kicking can lead
to the formation of a state which is localized near the kicking site. 
If $T$ is small enough, this is a bound state, while if $T$ is large, this is 
a resonance in the continuum of bulk states~\cite{bic,reyes,thuberg} as we 
will discuss. In 
Sec.~\ref{sec_cond}, we will see how a time-independent potential at one site
affects the transmission and how periodic kicking at that site can lead to 
an increase in the transmission. In Sec.~\ref{sec_inter}, we will extend
the model to include spin and will introduce a Hubbard like interaction
between spin-up and spin-down electrons at the same site which is subjected
to periodic kicks. We again study the effects of the interaction on the
transmission of a two-particle wave packet~\cite{dhar} which is in a spin 
singlet state. We will also study the possibility of bound states in this 
system. We will end in Sec.~\ref{sec_concl} with a summary of our results 
and some directions for future work.

\section{The model}
\label{sec_ham}

\begin{figure}
\includegraphics[width=0.98\columnwidth]{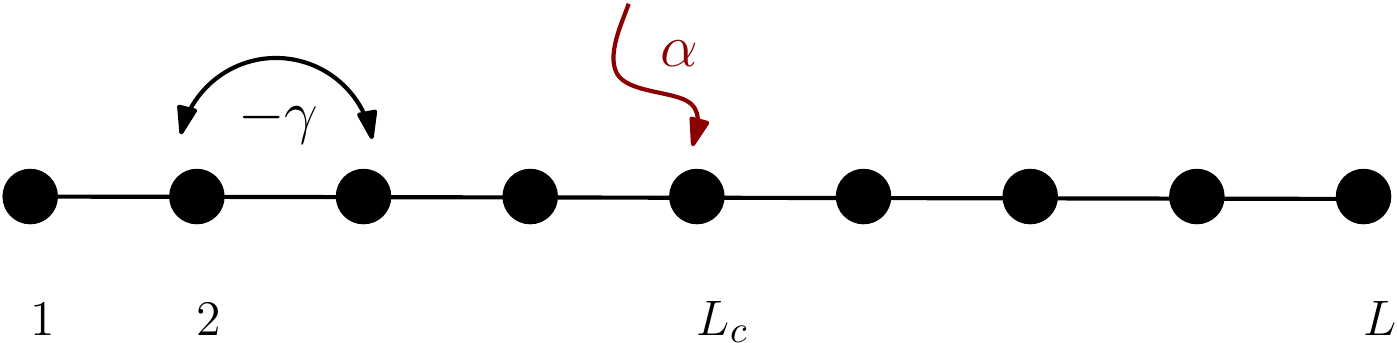}
\caption{Schematic figure of a one-dimensional lattice where a fermion can 
hop between nearest-neighbor sites with amplitude $-\ga$. A periodic
$\de$-function kick is applied at the central site $L_c$ of a lattice of 
length $L$. The kicking strength is $\al$.} 
\label{fig:schema} \end{figure}

We consider a chain of length $L$ on which spinless electrons hop between 
neighboring sites with the Hamiltonian
\beq H_{TB} ~=~ - ~\ga ~\sum_{n=1}^{L-1} ~(c^\dg_n c_{n+1} + H.c.), 
\label{htb} \eeq
where $\ga$ is the hopping integral, and $c^\dg_n$ and $c_n$ are the fermion 
creation and annihilation operators at site $n$ respectively. (We will set 
$\ga =1$ in all our numerical calculations. We will also set the 
lattice spacing and $\hbar$ to 1 in this paper).
The energy-momentum dispersion for this Hamiltonian is given by $E_k = - 2 
\ga \cos k$, where $k$ lies in the range $[-\pi,\pi]$; hence the group 
velocity is $v_k = |2 \ga \sin k|$. We now apply periodic 
$\de$-function kicks at a single site labeled as $L_c$ lying in the 
middle of the system; the kicks are described by the time-dependent potential
\beq H_K ~=~ \al ~\sum_{m=-\infty}^\infty ~\de(t-mT) ~c^{\dg}_{L_c}
c_{L_c}. \eeq 
Hence the complete Hamiltonian (see \Fig{fig:schema}) is 
\beq H = H_{TB} + H_K. \eeq
We are interested in studying the properties of this system as we tune 
parameters such as the strength $\al$ and the time period $T$ of the kicking.

\section{Wave packet dynamics and transport}
\label{sec_trans}

\begin{figure}
\begin{center}
\subfigure{\includegraphics[width=0.98\columnwidth]{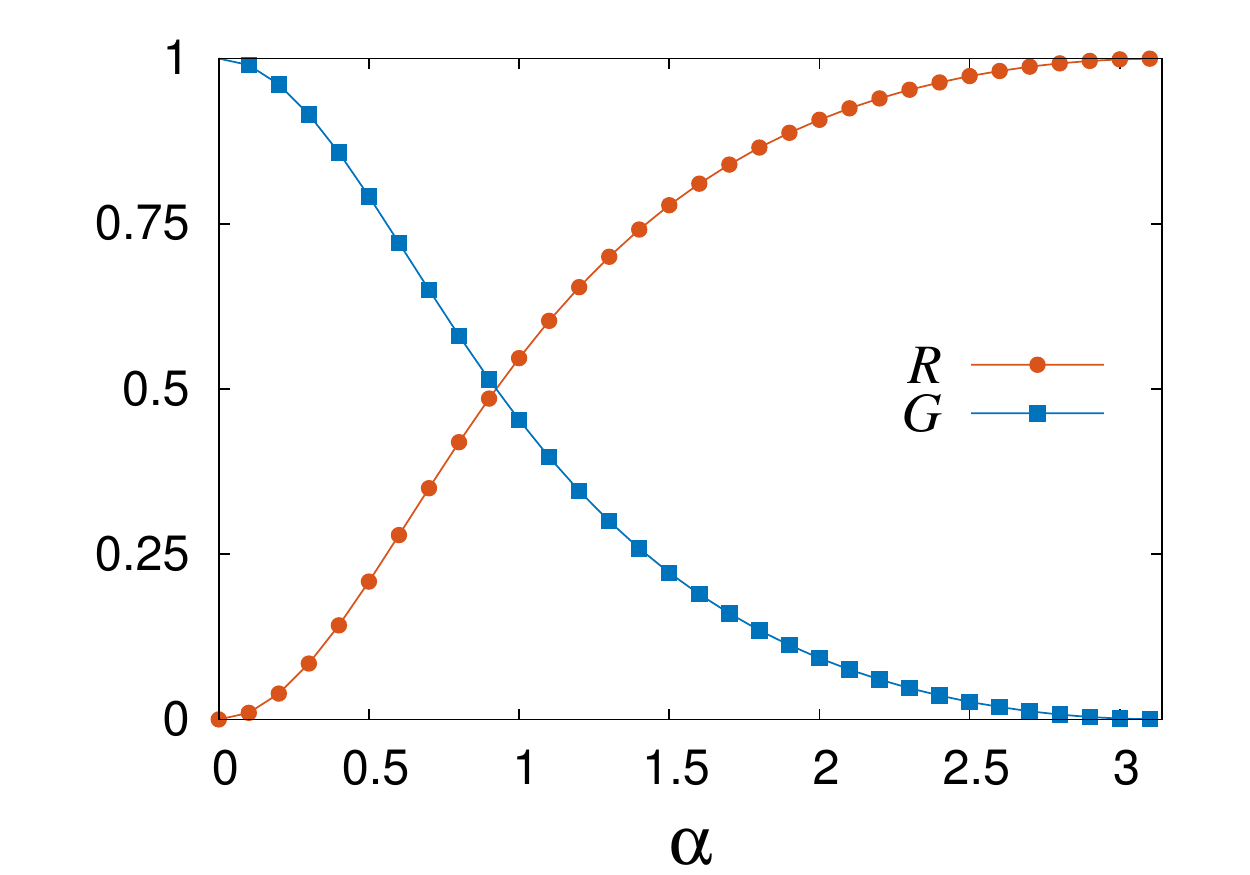}} \\
\subfigure{\includegraphics[width=0.98\columnwidth]{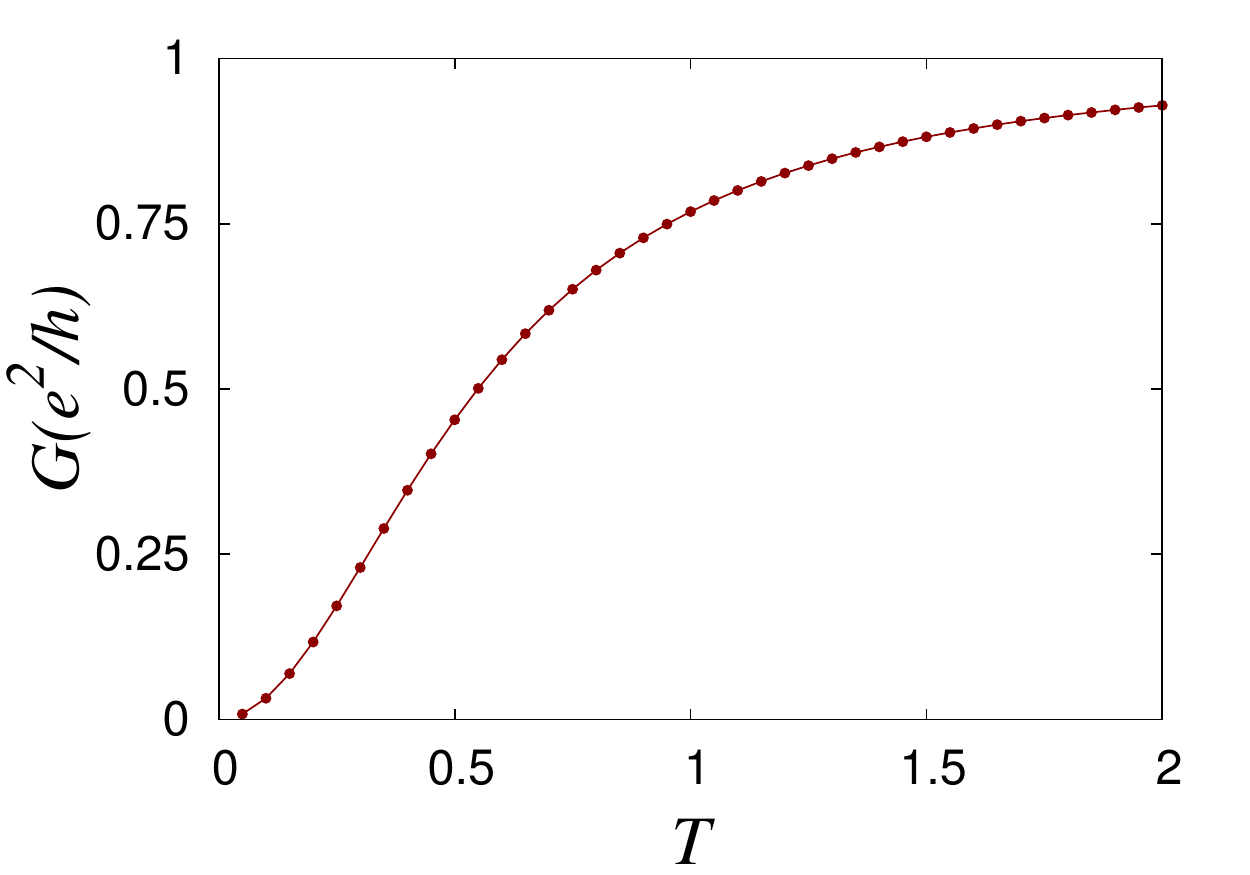}}
\end{center}
\caption{(Top) Reflection probability $\cal R$ (red circles) and different
conductance $G = (e^2/h) {\cal T}$ (blue squares) 
vs $\al$ of a wave packet which is centered at the site $L_o=50$ with width
$\si=5$ on a lattice with $L=401$ sites. The kicking is done at the $L_c=
201$-th site. The kicking time period is $T=0.5$, and the momentum is centered 
at $k_c = \pi/2$. The wave packet is evolved up to a time $(L-2L_o)/|2 \sin 
k_c|$. (Bottom) Differential conductance $G = (e^2/h) {\cal T}$ vs $T$ when 
$\al=1$ is kept fixed. Other parameters are the same as in the top panel.} 
\label{fig:trans} \end{figure}

\begin{figure}
\includegraphics[width=\columnwidth]{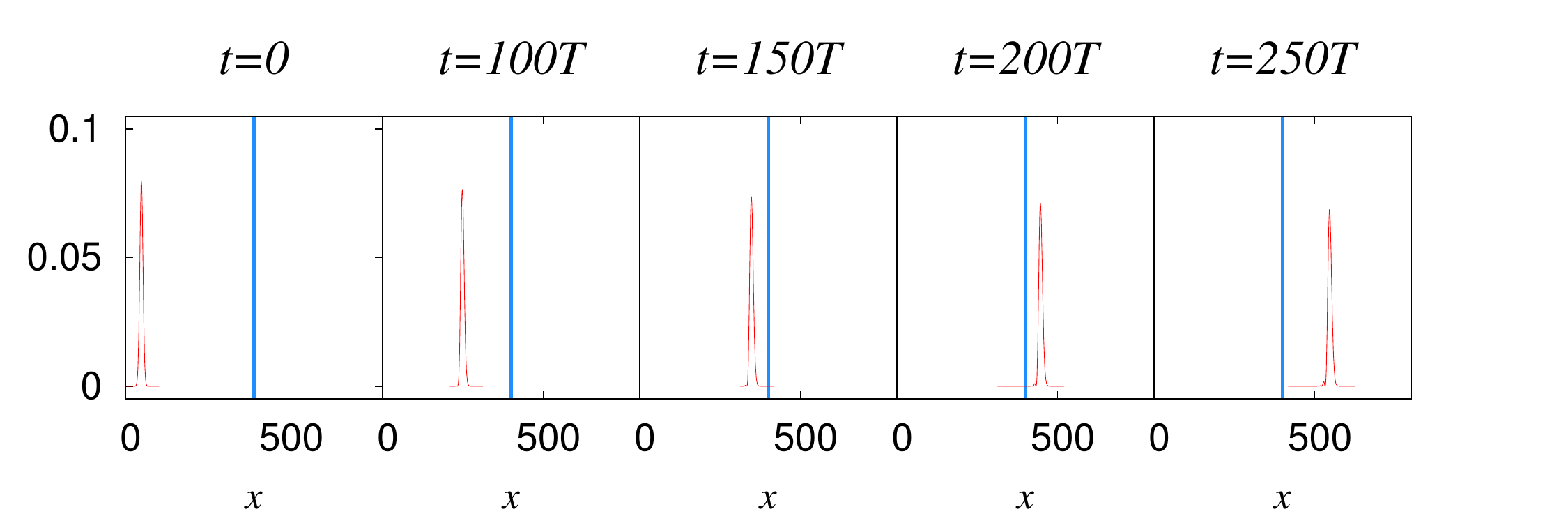}
\includegraphics[width=\columnwidth]{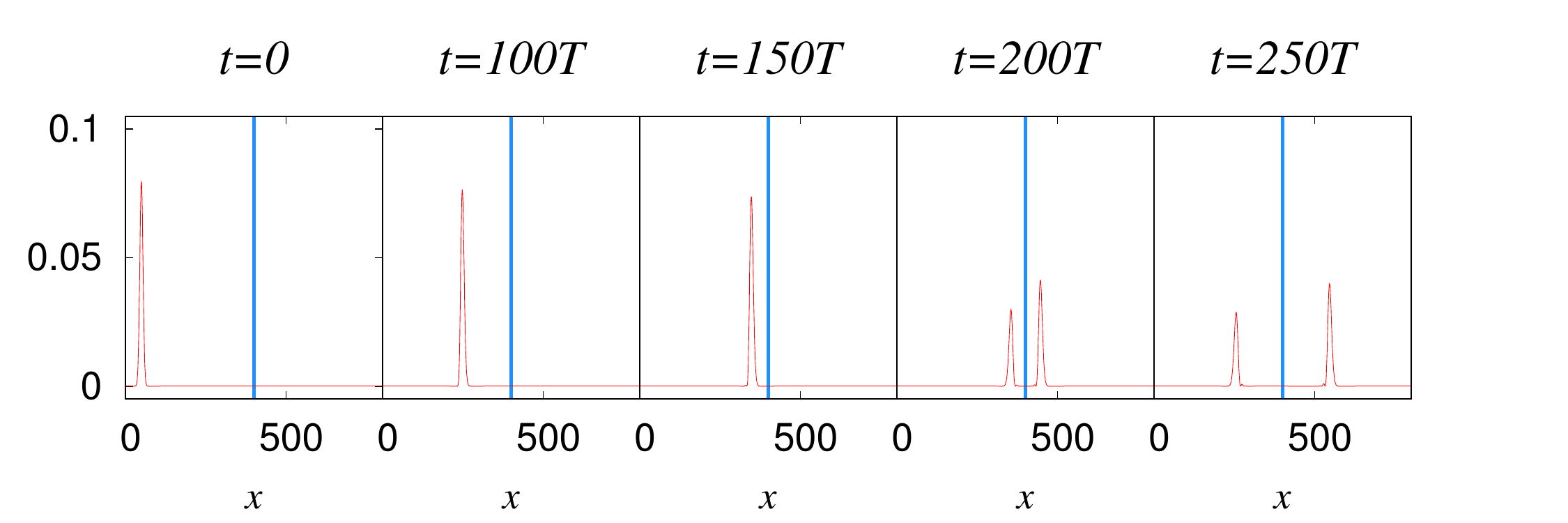}
\includegraphics[width=\columnwidth]{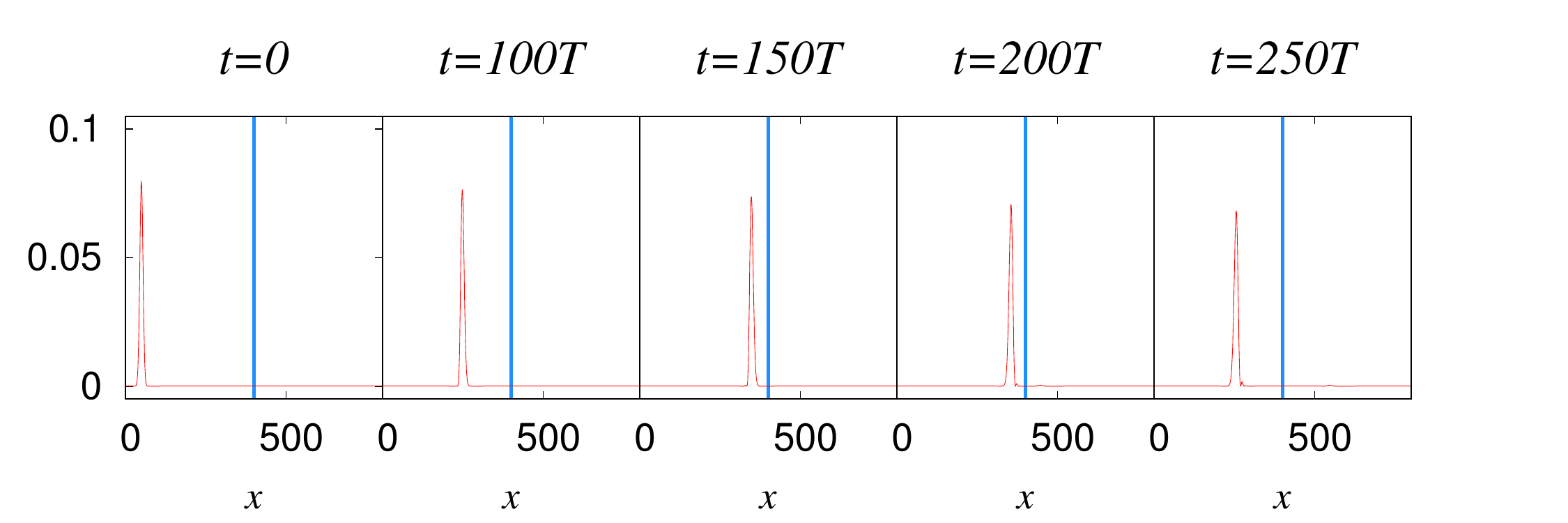}
\caption{Transmission and reflection of a wave packet 
which is centered at the site $L_o=50$ with width $\si=5$ on a lattice with
$L=801$ sites. The kicking is done at the $L_c=401$-th site (denoted by a 
vertical blue line). The kicking time period is $T=1$, and the momentum 
is centered at $k_c = \pi/2$. The wave packet is shown at various intervals
of time. From top to bottom, the different cases correspond to kicking 
strengths $\al = 0, ~1.4$ and $3$. In the first case, the wave packet 
moves across the kicked site unhindered. In the second case, the original 
wave packet splits into two, one which transmits across the barrier and the 
other which reflects. In the third case, the wave packet gets completely 
reflects from the central site.} \label{fig:transwave} \end{figure}

We will first investigate the effect of the kicking on the transport 
properties. To this end, we first construct an initial wave packet
at time $t=0$ given by
\beq \psi_i (r) ~=~ \frac{1}{(2\pi \si^2)^{1/4}} ~\exp \Big(-\frac{
(r-L_o)^2}{4 \si^2} ~+~ ik_c r \Big), \label{eqn:wavepa} \eeq
which satisfies $\int dr |\psi_i (r)|^2 = \sum_n |\psi_i (n)|^2 = 1$. 
Here $\si$ denotes the width of the wave packet in real space, $k_c$ is 
the central value of the wave vector of the wave packet, and $L_o$ is the 
position in real space where the wave packet is initially centered. Since 
the wave packet is centered at the momentum $k_c$ we know that the effective 
group velocity of the packet will be $|2\sin k_c|$. We evolve the system for 
a time $(L-2L_o)/|2\sin k_c|$; this allows the wave packet the time to travel
a distance $L/2-L_o$ when it reaches the site where the periodic kicks
are applied and then allows the transmitted part of the wave packet to travel 
further by an equal distance $L/2-L_o$. At the end of that time, we have
a wave function $\psi_f$; we then define the 
transmission and reflection probabilities $\cal T$ and $\cal R$ as 
\bea {\cal R} &=& \sum_{n=1}^{L_c} ~|\psi_f (n)|^2, \notag \\ 
{\cal T} &=& \sum_{n=L_c+1}^L ~|\psi_f (n)|^2. \label{rt} \eea
These definitions ensure that ${\cal R} + {\cal T}=1$.
(For spinless electrons, the transmission probability ${\cal T}$ at an energy 
$E$ is related to the zero temperature differential conductance $G = dI/dV$ as
$G (E)=(e^2/h) {\cal T}(E)$. In our figures, we will plot $G$ rather than 
$\cal T$, since $G$ is a directly measurable physical quantity).
The numerical results for $k_c = \pi/2$ are shown in \Fig{fig:trans}. 
(A reason for choosing $k_c = \pi/2$ is that this minimizes the rate of 
spreading of the wave packet~\cite{seshadri}. In one dimension, it is known 
that the width of a wave packet spreads in time at a rate which is 
proportional to $(\partial^2 E_k /\partial k^2)_{k=k_c} = 2 \ga \cos k_c$;
this vanishes at $k_x = \pi/2$). An important point to note in 
\Fig{fig:trans} is that $\cal R$ goes to 1 and $G = (e^2/h) {\cal T}$ 
goes to zero as $\al$ approaches $\pi$. Hence there is perfect reflection at a 
particular value of $\al$. This can be seen more clearly by directly observing 
the evolution of a wave packet in the presence of the periodic kicking. Some 
representative cases are shown in \Fig{fig:transwave}. Here $\cal T$ and 
$\cal R$ are calculated for a wave packet which is centered at the site 
$L_o=50$ with width $\si=5$ on a lattice with $L=801$ sites. The kicking is 
done at the $L_c=401$-th site (denoted by a vertical blue line). The kicking 
time period is taken to be $T=1$, and the central momentum of the wave packet 
is taken to be $k_c = \pi/2$. The wave packet is shown at different intervals 
of time. From top to bottom, the different cases correspond to kicking 
strengths $\al = 0, ~1.4$ and $3$. For the first case when there is no 
kicking, the wave packet moves 
across the kicked site unhindered. For the second case, one sees that the 
original wave packet splits into two, one which transmits across the barrier 
and the other which reflects. For the third case, when $\al$ is close to 
$\pi$, one finds that wave packet gets completely reflects from the central 
site.

It is also interesting to see what happens when $\al$ is fixed at a 
particular value and $T$ is varied. This is shown in \Fig{fig:trans}. 
We notice that at small $T$, the transmission is extremely small, a feature 
which we find to be generic in most cases for non-zero $\al$. 


%
%
%
%
%

\section{Perfect reflection and dynamical localization}
\label{sec_dynloc}

A curious feature noted in the last section is that the wave packet 
\emph{completely} reflects when the kicking strength $\al$ is close to 
$\pi$. This is intimately related to dynamical localization. We will make 
this connection clear in this section. It has been shown in previous 
work~\cite{agarwala16,agarwala17} that a periodic kicks of strength $\pi$ on 
one sublattice of a bipartite system can lead to the phenomena of dynamical 
localization where a wave packet remains localized in space; this holds even 
when there is no disorder present in the system.

In the present context, the time evolution operator for a single time 
period $T$ can be written as
\beq U ~=~ \exp(-i \al c^\dg_{L_c} c_{L_c}) ~\exp [i \ga T 
\sum_{n=1}^{L-1} (c^\dg_n c_{n+1} + H.c.)]. \label{ut} \eeq
It is particularly instructive to look at $U^2$ which evolves the system 
for a period $2T$. We rewrite $H_{TB}$ in Eq.~\eqref{htb} as 
\beq H_{TB} ~=~ H_r ~-~ \ga ~(c^\dg_{L_c}c_{L_c+1} + c^\dg_{L_c}
c_{L_c-1} + H.c.), \eeq 
where $H_r$ denote the rest of the terms. Then
\bea U^2 &=& e^{-i \al c^\dg_{L_c} c_{L_c}} ~\exp(-i H_{TB} T) ~
\notag \\
&& \times ~e^{-i \al c^\dg_{L_c} c_{L_c}} ~\exp(-i H_{TB} T). \eea
We can evaluate this for $\al = \pi$ by noting that $e^{-i \pi c^\dg_{L_c} 
c_{L_c}} = e^{i \pi c^\dg_{L_c} c_{L_c}}$ (since $c^\dg_{L_c} c_{L_c}$ can
only take the values 0 and 1), and using the identities
\bea e^{i \pi c^\dg_{L_c} c_{L_c}} c_{L_c} e^{-i \pi c^\dg_{L_c} c_{L_c}} 
&=& - ~c_{L_c}, \non \\
e^{i \pi c^\dg_{L_c} c_{L_c}} c^\dg_{L_c} e^{-i \pi c^\dg_{L_c} c_{L_c}} 
&=& - ~c^\dg_{L_c}. \eea
We then find 
\bea U^2 &=& \exp[-i T \{ H_r + \ga (c^\dg_{L_c}c_{L_c+1} + 
c^\dg_{L_c}c_{L_c-1} + H.c.) \} ] \notag \\ 
&\times& \exp[-i T \{ H_r - \ga (c^\dg_{L_c}c_{L_c+1} + c^\dg_{L_c}
c_{L_c-1} + H.c.) \}]. \notag \\
&& \label{u2} \eea
Using the Baker-Campbell-Hausdorff formula 
\beq e^X ~e^Y ~=~ e^{X + Y + \frac{1}{2} [X,Y] + \cdots}, \label{bch} \eeq
and assuming that $\ga T
\ll 1$, we can evaluate Eq.~\eqref{u2} to first order in $T$; we obtain 
\beq U^2 ~=~ \exp(-i2H_r T). \eeq
We now examine the form of $H_r$. We see that $H_r$ is the part of the 
tight-binding Hamiltonian in which the hoppings to the central site are 
removed, i.e., $H_r$ is effectively described by two disconnected chains. This 
is the underlying reason why a wave packet completely reflects back at 
$\al = \pi$. Interestingly, this is also the regime which leads to dynamical 
localization in translationally invariant systems where the periodic kicking 
is applied to all the sites on one sublattice of a bipartite 
lattice~\cite{agarwala16,agarwala17}. 
 
We note here that the parameter $\al$ appearing in Eq.~\eqref{ut} is
really a periodic variable, namely, $\al$ and $\al + 2 \pi$ give the same 
results since $c^\dg_{L_c} c_{L_c}$ can only take the values 0 and 1.
In particular, $\al$ equal to any integer multiple of $2\pi$ will have 
no effect on the time evolution.

For later purposes, it is convenient to consider the Floquet eigenstates 
$\psi_j$ and eigenvalues $e^{-i \ep_j T}$ of the unitary operator $U$ defined 
in Eq.~\eqref{ut}. The $\ep_j$'s are called quasienergies; since they are only
defined modulo $2\pi/T$, we can take them to lie in the range $[-\pi/T,\pi/T]$.

\section{Non-Equilibrium Green's function method}
\label{sec_negf}

The non-equilibrium Green's function (NEGF) method is one of the most robust 
methods for evaluating the conductance of a time-independent 
Hamiltonian~\cite{datta}. Here, we extend it to a periodically driven system 
and show that such a formalism appears to work for large driving frequencies 
or small time periods $T$.

The time evolution operator for a single time period $T$ can be written as
\bea U &=& \exp(-i \al c^\dg_{L_c} c_{L_c}) ~\exp [i \ga T \sum_{n=1}^{L-1} 
(c^\dg_n c_{n+1} + H.c.)] \notag \\ 
&\equiv& \exp(-i H_{eff} T), \label{heff} \eea
where $H_{eff}$ can be found exactly by a numerical calculation. We now 
propose to use $H_{eff}$ as a time-independent Hamiltonian and implement the 
NEGF method. Namely, we use the Hamiltonian $H_{eff}$, along with the 
self-energies 
$\Sigma_1 (E_c)$ and $\Sigma_2 (E_c)$ at the left and right ends of the
system (here $E_c = - 2 \ga \cos k_c$ is the energy of a particle with
momentum $k_c$), to compute the zero temperature differential conductance $G$ 
at the energy $E_c$. (See Ref.~\onlinecite{agarwal06} for details of the 
procedure).

The comparison of the differential conductance $G (E_c)$ obtained 
using the NEGF method and the exact value using wave packet dynamics is 
shown in \Fig{fig:negf}. (An analytical expression for $G (E_c)$ will
be presented in Eq.~\eqref{trans} below for the case when $T$ is small).
It is clear that the NEGF method using $H_{eff}$
works well for small $T$, but deviates significantly as $T$ becomes large. 
It is natural to ask what determines the crossover time scale between the 
two regimes. Another observation from \Fig{fig:negf} is that, even when 
$\al \sim \pi$, the wave packet dynamics shows that the transmission 
$\cal T$ is quite far from zero when the time period $T$ is large. 
Both of these observations can be understood by the following argument. Since 
a wave packet with width $\si$ and centered at a momentum $k_c$ has a velocity
$|2\ga \sin k_c|$, it will take a time $\Delta t =\si /|2\ga \sin k_c|$ 
to cross any particular site on the lattice. If the kicking time period $T$ is 
larger than this $\Delta t$, one expects that the wave packet may not sample 
the kick and will therefore pass right through the site where the kicking is 
being applied. Therefore the kicking can properly affect the transmission only 
when 
\beq T ~\lesssim~ \frac{\si}{|2\ga \sin k_c|}. \label{cond0} \eeq
In \Fig{fig:negf}, we have chosen $\si = 5$ and $k_c = \pi/2$; this gives 
$T \lesssim 2.5$ in Eq.~\eqref{cond0}. This explains why the NEGF results 
agree well with those based on wave packet dynamics if $T = 0.4, ~1.2$ and 
$2.4$, but not if $T=3.6$.

We note that the use of an effective Hamiltonian is only justified if 
$2\ga T < \pi$; this can be seen as follows. We recall that the quasienergies 
$\ep_j$ are only defined up to multiples of the driving frequency $\om = 
2\pi /T$. Since the $\ep_j$'s are eigenvalues of $H_{eff}$, this means that 
$H_{eff}$ is not uniquely defined to begin with. The eigenstates of $H_{TB}$ 
in Eq.~\eqref{htb} lie in the range $[-2\ga,2\ga]$; hence if $2\ga T 
< \pi$, we can define the quasienergies of all the bulk states to lie in the 
range $[-\pi/T,\pi/T]$. This will define $H_{eff}$ uniquely.
On the other hand, the correspondence between the NEGF results and wave 
packet dynamics are expected to hold if the condition in Eq.~\eqref{cond0}
holds; this condition depends on both the wave packet width $\si$ and
the momentum $k_c$.

\begin{widetext}
\begin{center}
\begin{figure}
\begin{center}
\subfigure{\includegraphics[width=0.48\columnwidth]{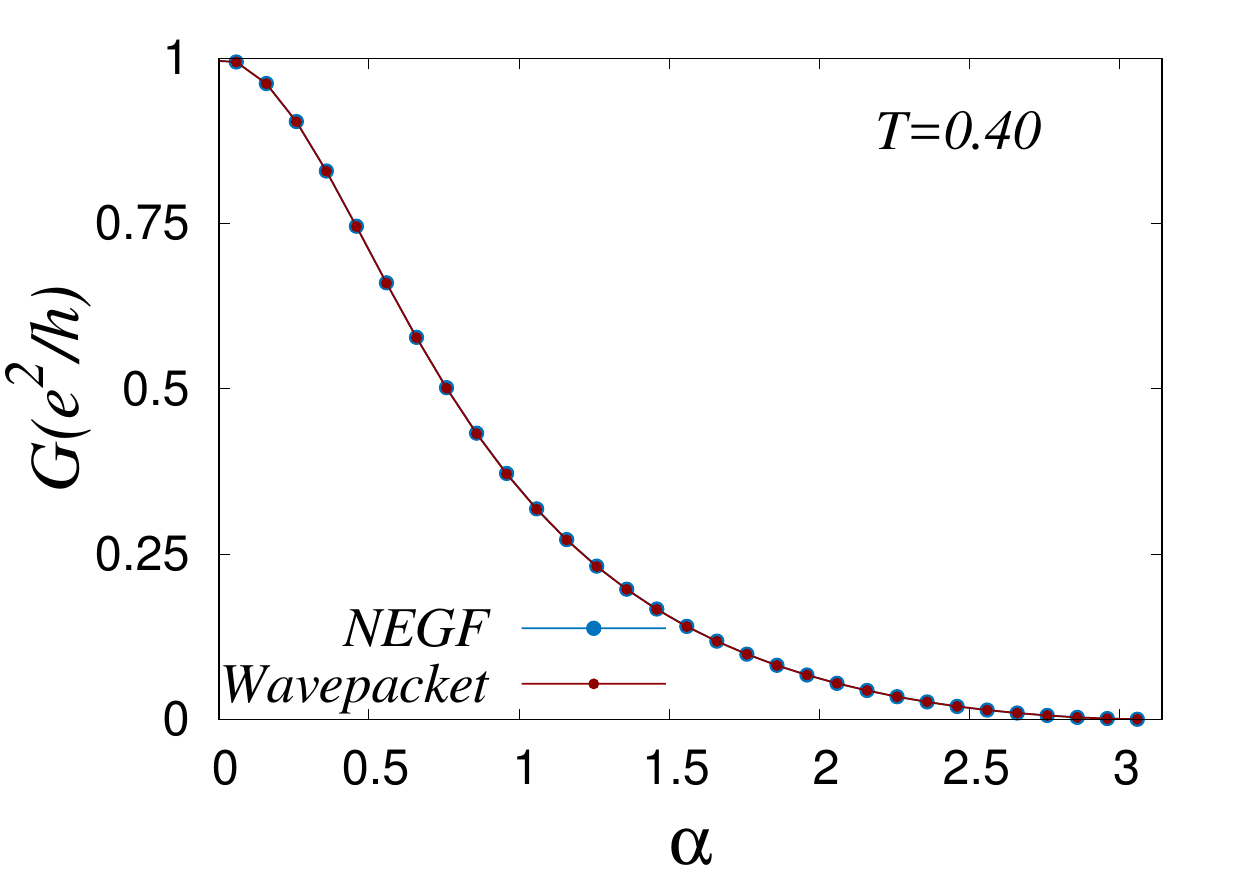}}
\subfigure{\includegraphics[width=0.48\columnwidth]{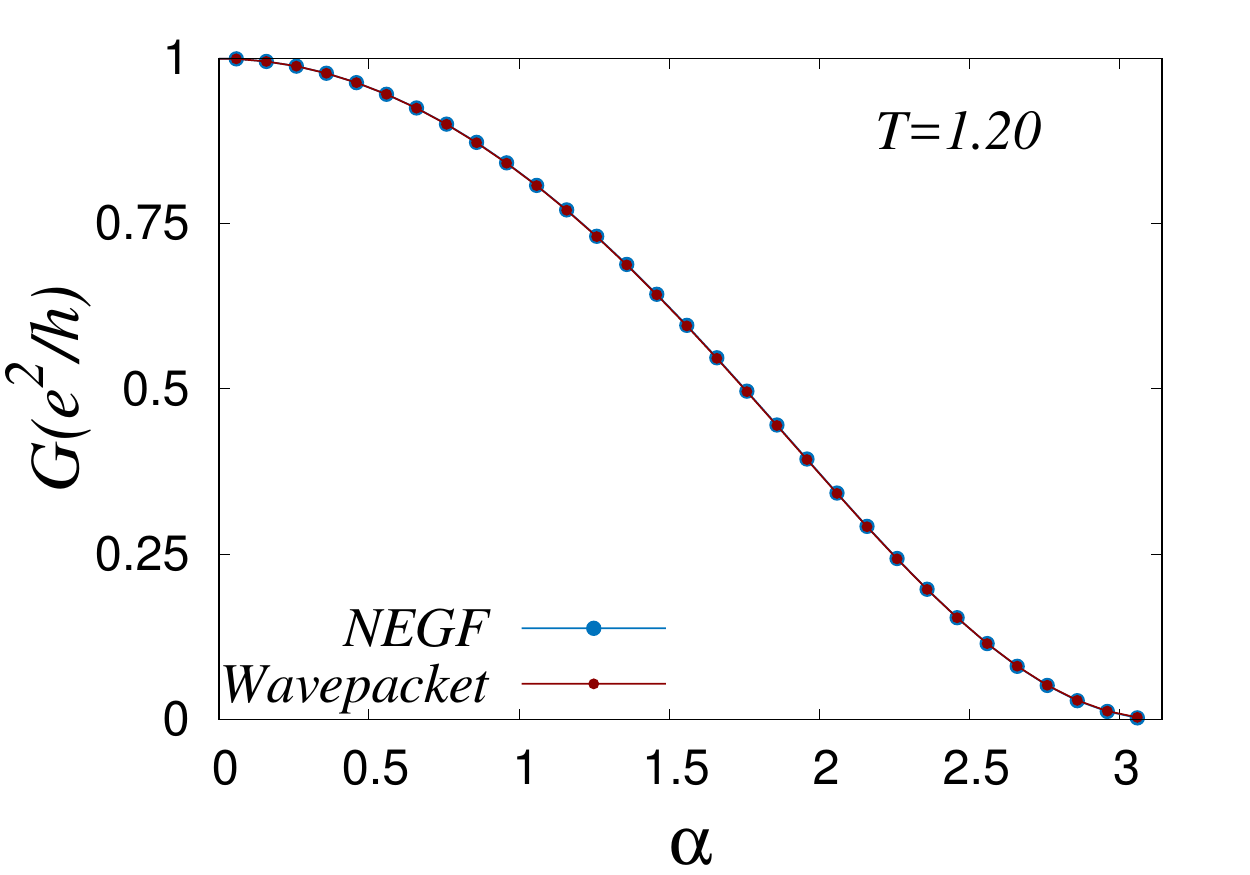}}
\subfigure{\includegraphics[width=0.48\columnwidth]{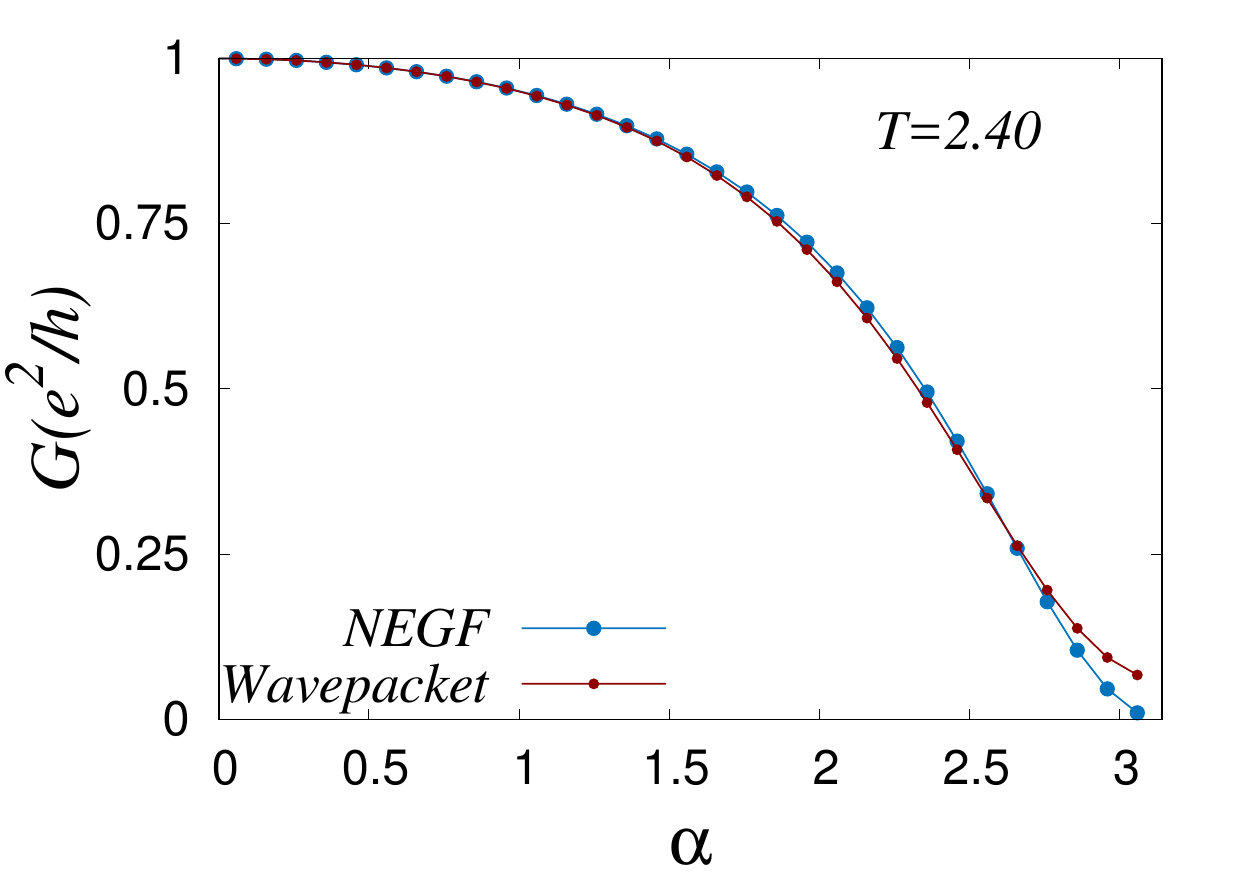}}
\subfigure{\includegraphics[width=0.48\columnwidth]{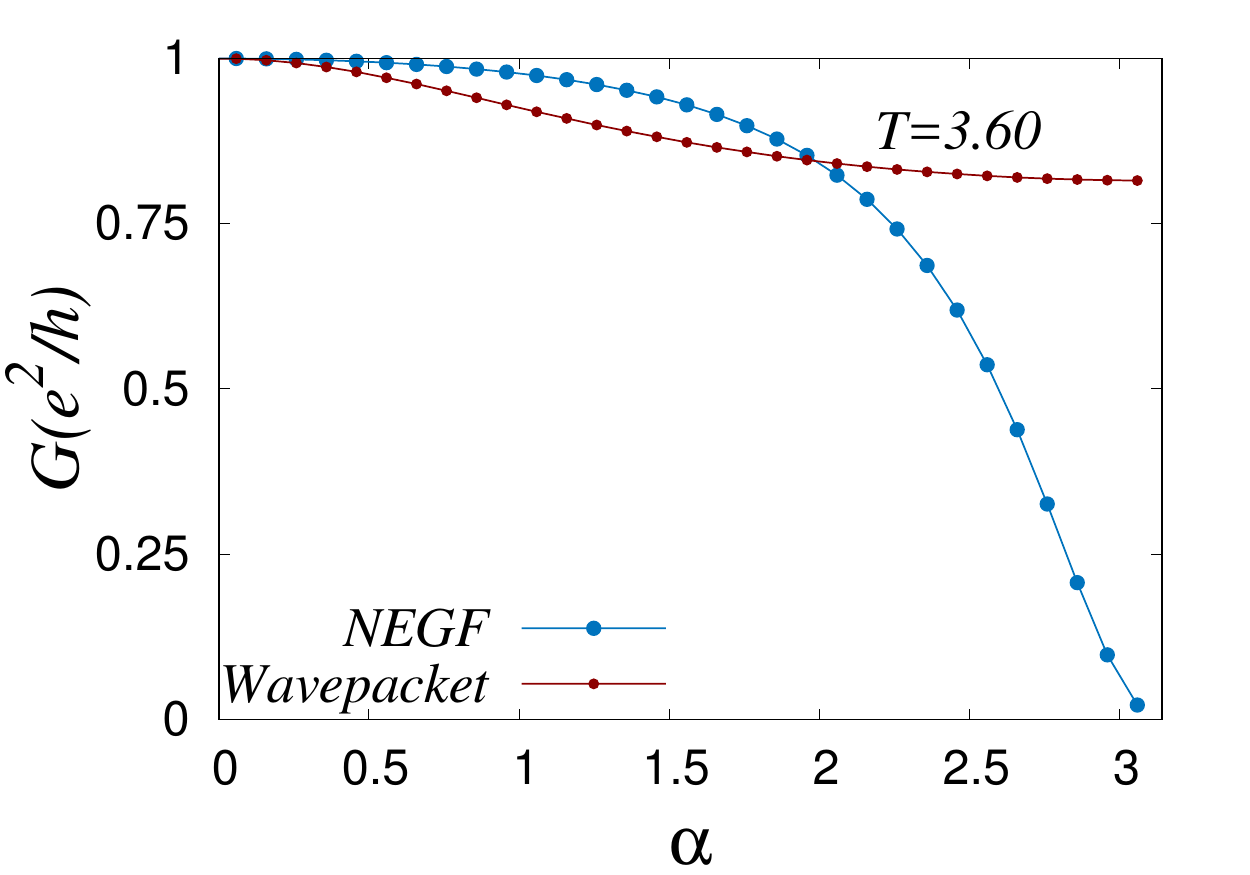}}
\end{center}
\caption{Differential conductance $G (E_c) = (e^2/h) {\cal T} (E_c)$ vs $\al$.
$G$ is computed from the dynamics of a wave packet which is centered at the 
site $L_o=50$ with width $\si=10$ on a lattice with $L=401$ sites. The kicking
is done at the $201$-th site, and the time period takes the values $0.4, 1.2, 
2.4$ and $3.6$ in the four figures from top left to bottom right. The momentum
is centered at $k_c = \pi/2$, so that $E_c = 0$.  The transmissions obtained 
from the exact wave packet dynamics and the NEGF formalism are compared. We 
see that the NEGF formalism matches the exact results for small $T$.} 
\label{fig:negf} \end{figure}
\end{center}
\end{widetext}

\section{Floquet bound states and resonances}
\label{sec_bound}

In the presence of kicking we can study if there are Floquet bound states in 
the system and explore the properties of such bound states both 
analytically and using numerical techniques.

At high frequencies (i.e., small values of $T$), the effective Hamiltonian
prescription, as briefly discussed in Sec.~\ref{sec_negf}, becomes more and 
more accurate. If both $\al$ and $\ga T$ are small, we can use Eq.~\eqref{bch}
to show that the effective Hamiltonian is 
\beq H_{eff} ~=~ H_{TB} ~+~ \frac{\al}{T} c^\dg_{L_c}c_{L_c} \eeq
to lowest order in $\al$ and $\ga T$. This is effectively a time-independent 
system with 
a potential equal to $\al/T$ at the site $L_c$. It is known that such a 
potential on a lattice gives a transmission probability 
\beq {\cal T} (k_c) ~=~ \frac{4 \ga^2 \sin^2 k_c}{4 \ga^2 \sin^2 k_c ~+~ 
\frac{\al^2}{T^2}} \label{trans} \eeq
for a particle which is coming in with momentum $k_c$ and energy $E_c$. The 
form in Eq.~\eqref{trans} explains the shape of the first plot in 
Fig.~\ref{fig:negf} where $T= 0.4$ is small. A potential $\al/T$ at one 
site also produces a bound state with energy $\ep_b$ given by
\beq \ep_b ~=~ \pm ~\sqrt{4 ~+~ \frac{\al^2}{T^2}}, \label{bounden} \eeq
where the sign of $\ep_b$ is the same as the sign of $\al/T$.

Numerically, given all the eigenstates of either a time-independent 
Hamiltonian $H$ or a time evolution operator $U$, the bound states can be 
identified quickly by looking at the values of the inverse participation 
ratio (IPR) of all the states. The IPR of a state $|\psi \rangle = \sum_{n=1}^L
\psi(n) |n\rangle$ is defined as $\sum_{n=1}^L |\psi (n)|^4$. 
Typically, states which are spread over the entire system of length $L$ have 
an IPR of the order of $1/L$, while a bound state with a decay length $\lam$
which is much smaller than $L$ will have an IPR of the order of $1/\lam$ 
which is much larger than $1/L$. Hence a plot of the IPR versus
the eigenstate number will clearly show the bound states~\cite{manisha13}.

The bound state with the energy given in Eq.~\eqref{bounden} has an 
exponentially decaying wave function of the form
\bea \psi (n) &=& {\cal N} \exp (- |n - L_c|/\lam) ~~~{\rm if} 
~~~\frac{\al}{T} ~<~ 0, \non \\
&=& {\cal N} ~(-1)^n \exp (- |n - L_c|/\lam) ~~~{\rm if} 
~~~\frac{\al}{T} ~>~ 0, \non \\
&& \label{wavefn} \eea
where the normalization constant ${\cal N} = \sqrt{\tanh (1/\lam)}$, and
the decay length $\lam$ is given by
\beq \lam ~=~ \left( {\rm arccosh} \sqrt{1 + \frac{\al^2}{4T^2}} 
\right)^{-1}. \label{decay} \eeq
If $\lam \gg 1$, one can show that the Fourier transform of the
wave function in Eq.~\eqref{wavefn} will have a peak at $k=0$ if $\al/T < 0$
and at $k = \pm \pi$ if $\al/T > 0$. (The Fourier transform of a wave function 
$\psi(n)$ is defined as ${\tilde \psi} (k) = \frac{1}{\sqrt{L}} \sum_{n=1}^L 
\psi (n) e^{-ikn}$).
The IPR of the wave function in Eq.~\eqref{wavefn} is given by
\beq {\rm IPR} ~=~ \frac{\al}{T} ~\frac{\frac{\al^2}{T^2} + 2}{\Big(\frac{
\al^2}{T^2} + 4\Big)^{3/2}}. \label{boundIPR} \eeq

\begin{figure}
\begin{center}
\subfigure{\includegraphics[width=\columnwidth]{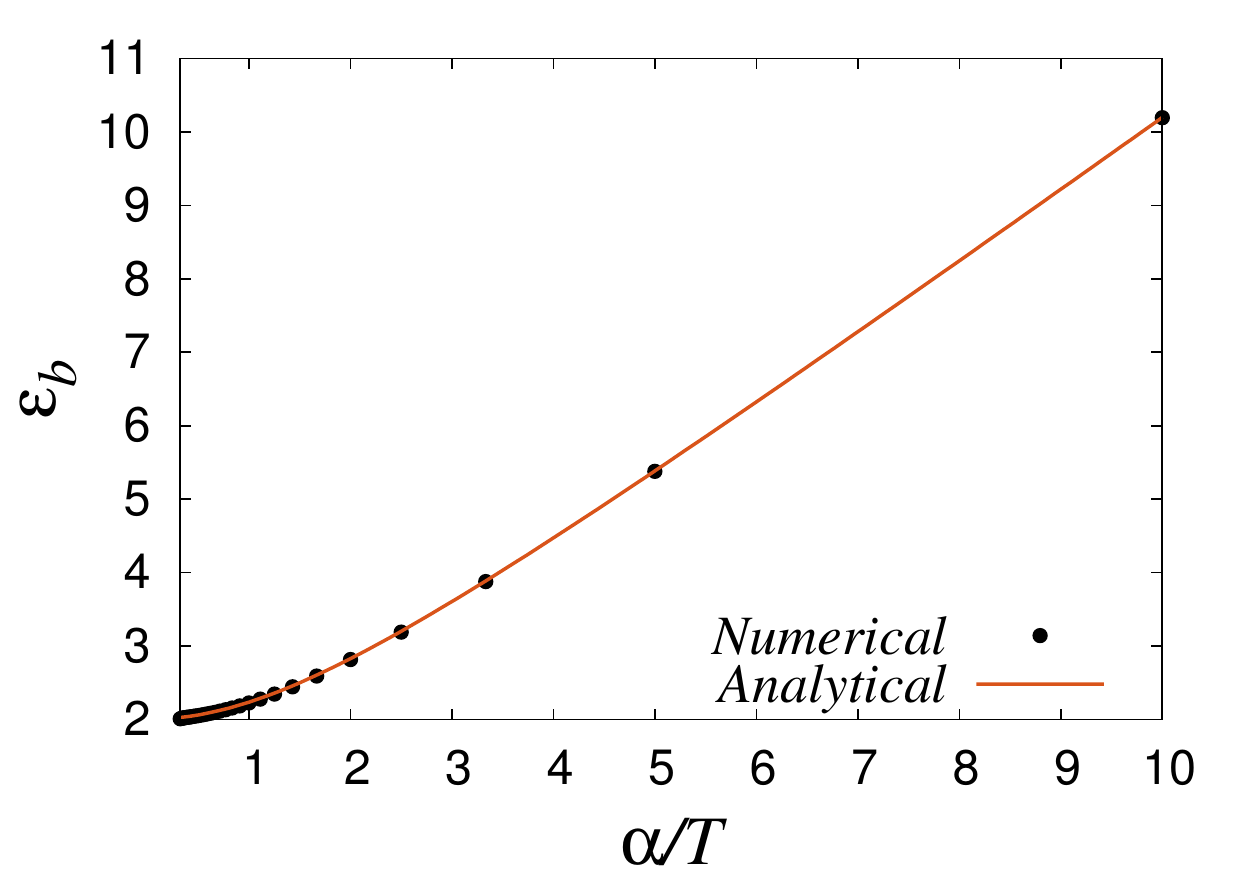}}
\subfigure{\includegraphics[width=\columnwidth]{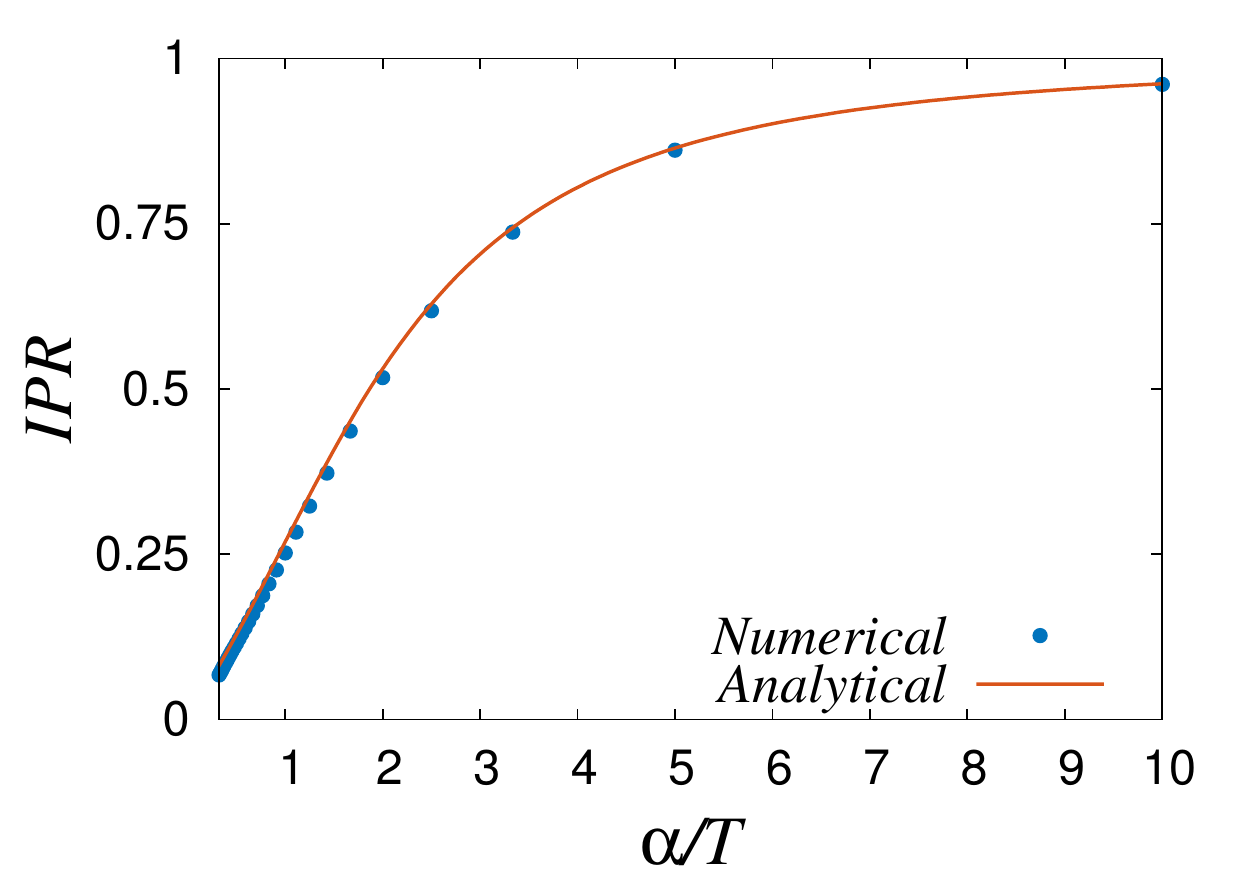}} 
\end{center}
\caption{Bound state energy and IPR for $\al=0.5$ and different values of $T$.
The numerically calculated spectrum matches well the analytical expression as 
shown in Eqs.~\eqref{bounden} and \eqref{boundIPR}, for $L=401$ and $L_c=201$.}
\label{figanaly} \end{figure}

The highest IPR and its corresponding quasienergy calculated numerically for
the eigenstates of the time evolution operator $U$ in Eq.~\eqref{heff} and 
their comparison with the analytical expressions in Eqs.~\eqref{bounden} and 
\eqref{boundIPR} is shown in \Fig{figanaly}. We will see later that the highest
IPR corresponds to a bound state in certain regions of the ``phase diagram" in
the $\al-T$ plane but to a resonance in the continuum in other regions.

It is interesting to study the full phase diagram for this system. This is 
shown in the left panel of \Fig{fig:IPRphase} when there is no time-independent 
on-site potential (i.e., $V=0$, where $V$ is defined in Eq.~\eqref{hv}). With 
increasing $\al$ one finds 
that the IPR increases, while increasing $T$ reduces IPR. Both of these are 
expected results since the effective potential due to the kicking is given by 
$\al/T$. However we find that the bound state appears to vanish abruptly 
when $T$ increases beyond $\pi/2$. This value of $T$ corresponds to the
driving frequency $\om = 4$ which is also the band width of the tight-binding 
model with $\ga =1$. Since the quasienergies of the bulk states (namely, the 
states which are extended throughout the system) form a continuum going from 
$-2\ga$ to $2 \ga$. Hence, for $T < \pi/2$, the quasienergies do not cover the
full range $[-\pi/T,\pi/T]$; this makes it possible for a bound state to appear
with a quasienergy which does not lie in the range of the bulk quasienergies;
hence the bound and bulk states do not mix. However,
for $T > \pi/2$, the bulk quasienergies cover the full range; hence any 
bound states must have a quasienergy which lies in the continuum of the bulk 
quasienergies. Such a situation is generally not possible except in special 
cases where the bound and bulk states cannot mix due to some symmetry or 
topological reasons; see Ref.~\onlinecite{bic} and references therein. 
Thus the disappearance of bound states above a certain value of $T$ is a 
unique feature of the Floquet system, since in a time-independent system in 
one dimension, a non-zero potential will always produce a bound state. 
Although there are no bound states for $T > \pi/2$, we will now see that there 
can be a resonance in the continuum; such a state is a superposition of a 
state which is localized near one point and some of the bulk states.

\begin{widetext}
\begin{center}
\begin{figure}
\begin{center}
\subfigure{\includegraphics[width=0.48\columnwidth]{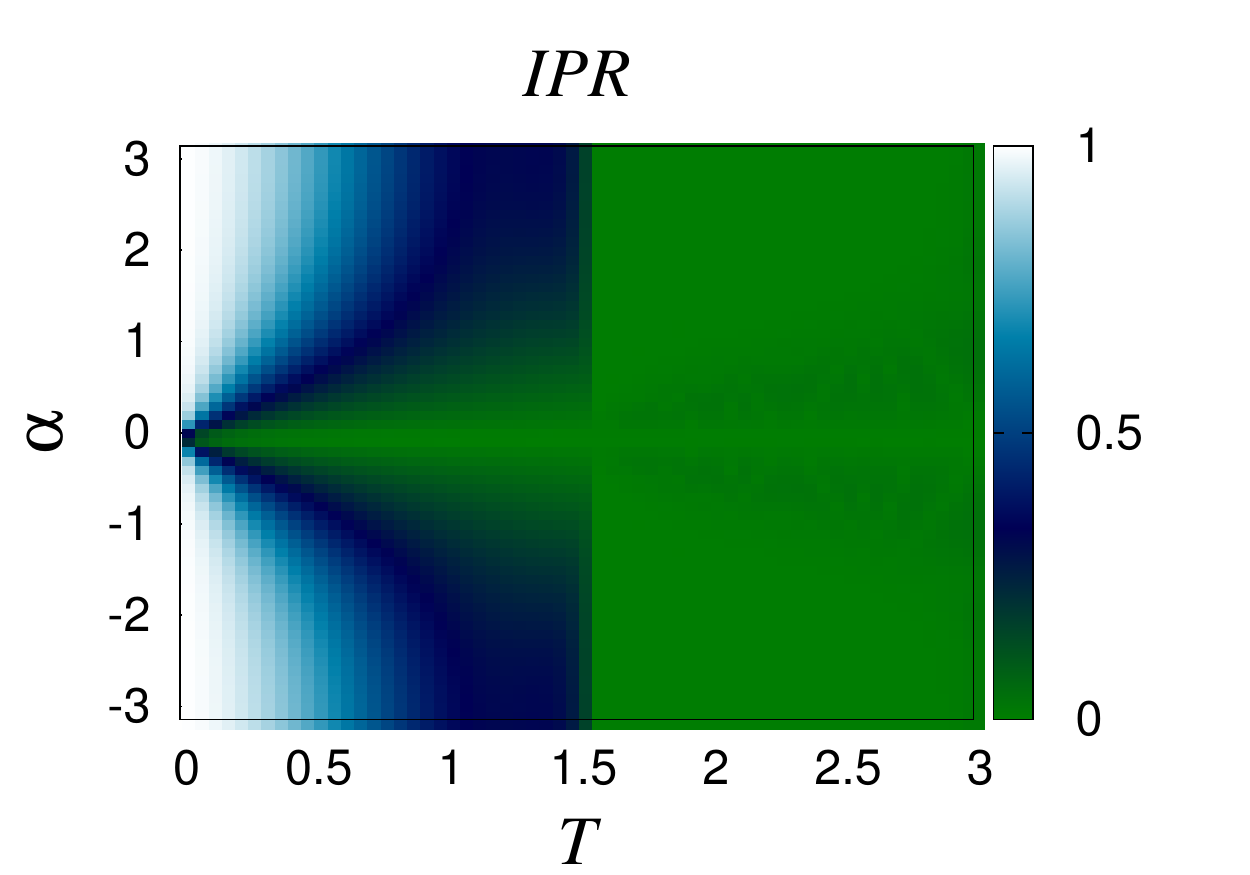}} 
\subfigure{\includegraphics[width=0.48\columnwidth]{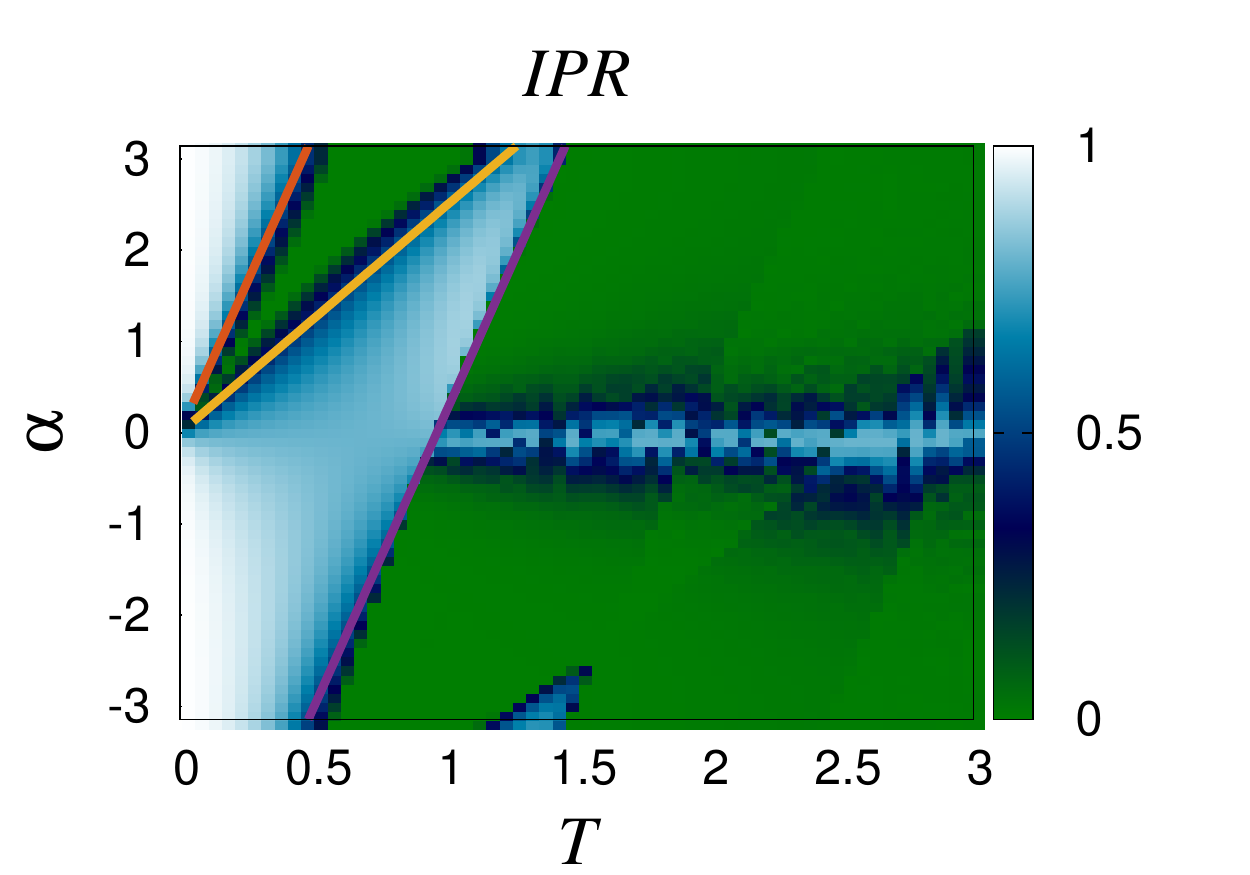}}
\end{center}
\caption{
The maximum IPR value of the eigenstates of the time evolution operator as 
a function of $T$ and $\al$ for $V=0$ (left panel) and $V=-4$ (right panel). 
The system has $L=401$ sites and the central site is kicked periodically. For 
$V=0$, the IPR increases as $\al$ increases, while increasing $T$ reduces the 
IPR. When $T$ crosses $\pi/2$ the bound state (which has the largest IPR) 
ceases to exist.} \mylabel{fig:IPRphase} \end{figure}
\end{center}
\end{widetext}

In Fig.~\ref{fig:bound}, we show the Floquet eigenvalues (since the time 
evolution operator is unitary, its eigenvalues lie on a unit circle in the 
complex plane), the probabilities $|\psi(n)|^2$ at different sites of a bound 
state, and the square of the modulus of the
Fourier transform of the bound state for a system with 401 sites in which
periodic $\de$-function kicks are applied at the $201$-th site with strength 
$\al=0.4$ and time period $T=1$. The bound state is easily identified because
it has the largest IPR equal to $0.0936$. Its Floquet eigenvalue is equal to
$-0.4473 - 0.8944i$ which is shown by a large red dot lying just outside the
continuum of the eigenvalues of the bulk states; this eigenvalue agrees
well with $\exp (-i \ep_b T) = -0.4518 - 0.8921i$, where $\ep_b = 
\sqrt{4+ \al^2/T^2}$ is the bound state energy given in Eq.~\eqref{bounden}.
According to Eq.~\eqref{decay}, the decay length of this state is equal to 
$\lam = 5$. The IPR equal to $0.0936$ agrees fairly well with the value of 
$0.1018$ given by Eq.~\eqref{boundIPR}. The square of the modulus of the 
Fourier transform, $|{\tilde \psi}(k)|^2$, of the bound state is found to 
have peaks at $k=\pm \pi$. 

\begin{widetext}
\begin{center}
\begin{figure}
\begin{center}
\subfigure{\includegraphics[width=0.32\columnwidth]{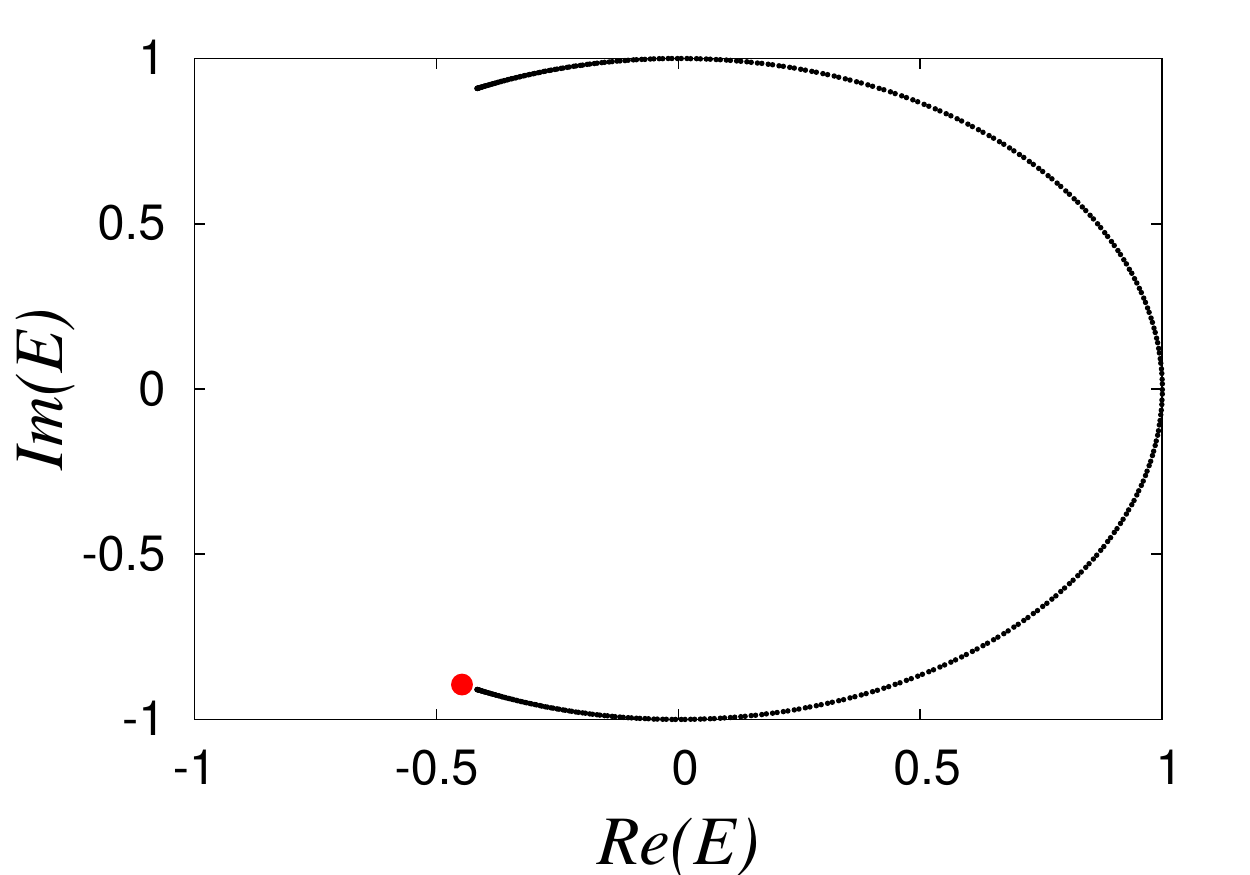}}
\subfigure{\includegraphics[width=0.32\columnwidth]{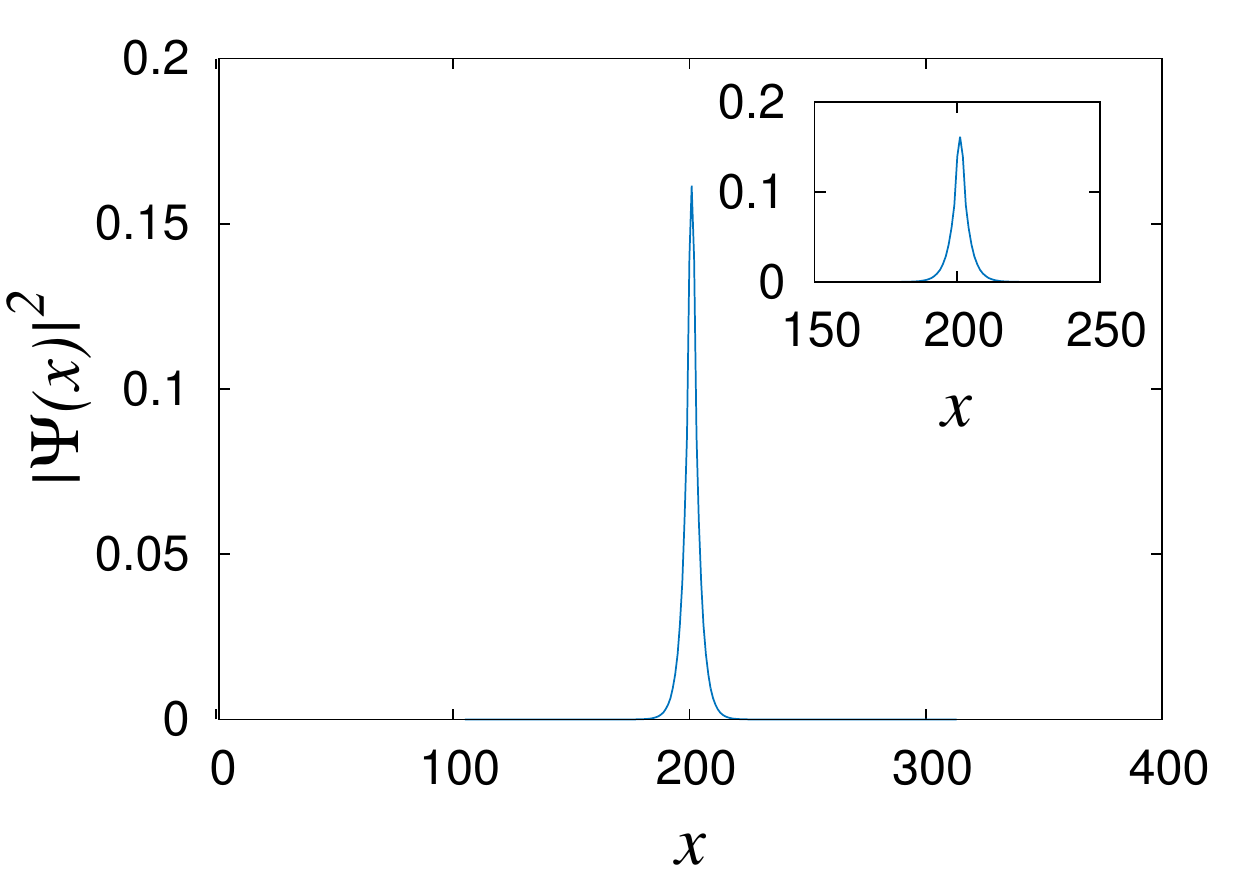}}
\subfigure{\includegraphics[width=0.32\columnwidth]{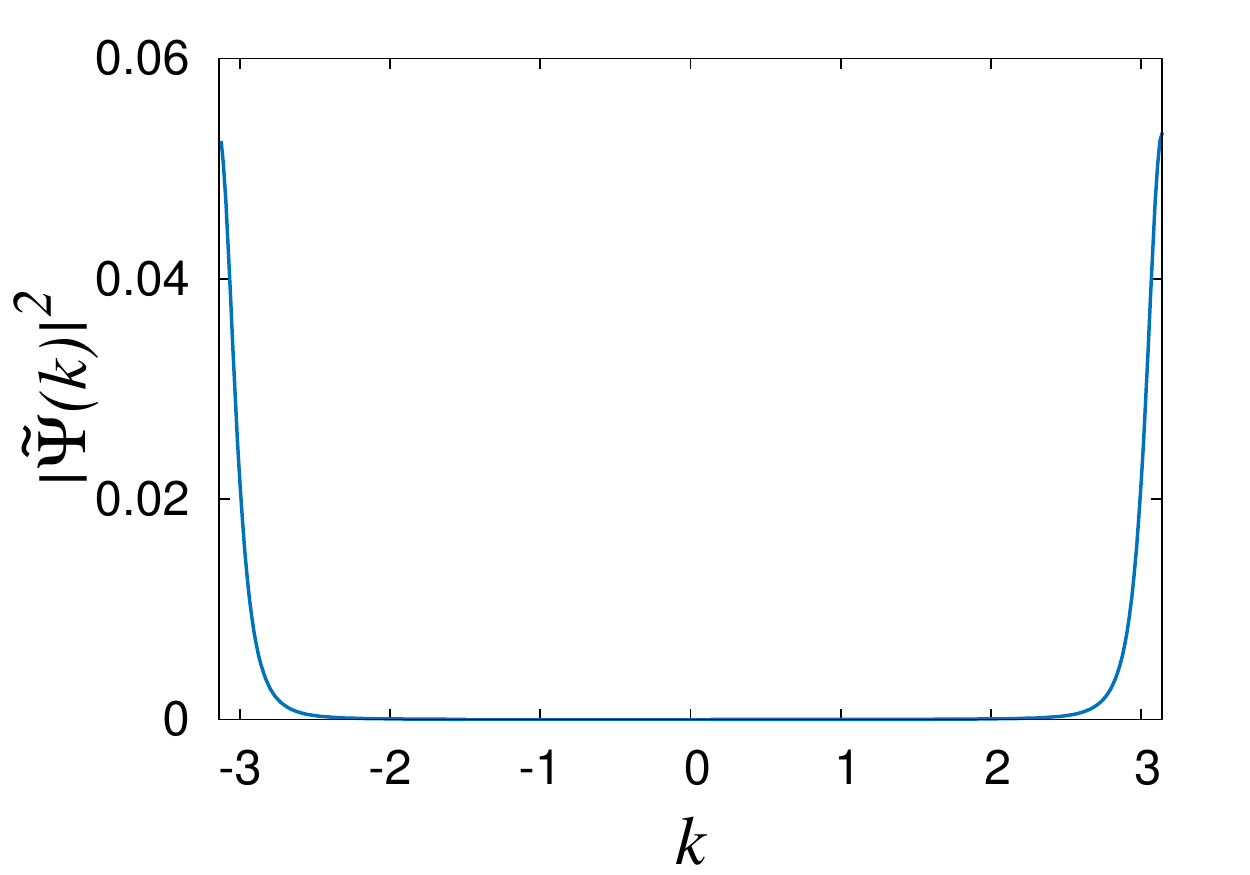}}
\end{center}
\caption{(Left) Eigenvalues of time evolution operator for a system with 401 
sites in which $\de$-function kicks are applied at the $201$-th site with 
$\al =0.4$ and $T=1$. There is a bound state with IPR equal to $0.0936$ and
Floquet eigenvalue equal to $-0.4473 - 0.8944i$ shown by a large red dot. 
(Middle) Probability $|\psi(n)|^2$ of the bound state. (Right) Square of the
modulus of the Fourier transform, $|{\tilde \psi}(k)|^2$, of the bound state. 
It has peaks at $k=\pm \pi$.} \label{fig:bound} \end{figure} 
\end{center}
\end{widetext}

Figure~\ref{fig:resonance} shows the Floquet eigenvalues, the probabilities 
$|\psi(n)|^2$ at different sites of a resonance state, and the square of the 
modulus of the Fourier transform of the resonance state for a system with 401 
sites in which $\de$-function kicks are applied at the $201$-th site with 
$\al=0.4$ and $T=2$. The resonance state has the largest IPR equal to $0.0324$.
Its Floquet eigenvalue is equal to $-0.6388 + 0.7694i$ which is shown by a 
large red dot lying within the continuum of the bulk eigenvalues; this value 
again agrees well with $\exp (-i \ep_b T) = -0.6384 + 0.7697i$, where $\ep_b$
is the bound state energy given in Eq.~\eqref{bounden} (the bound state has 
turned into a resonance here due to mixing with the bulk states). According to 
Eq.~\eqref{decay}, the decay length of this state is equal to $\lam = 10$. 
The IPR equal to $0.0324$ is significantly smaller than the value of $0.0502$ 
given by Eq.~\eqref{boundIPR}; this is because of a substantial mixing with 
plane waves with $k =\pm 0.967$ (found from the peaks in the Fourier 
transform). According to Eq.~\eqref{decay}, the decay 
length of this state is equal to $\lam = 10$. The square of the modulus of 
the Fourier transform, $|{\tilde \psi}(k)|^2$, of the bound state is found to 
have peaks at both $k=\pm \pi$ and $k= \pm 0.967$. We can understand the
peaks at $k= \pm 0.967$ as follows: we note that there are bulk states at 
these values of $k$ with a Floquet eigenvalue equal to $\exp (i 2 \ga T \cos k)
= -0.6444 + 0.7647i$. This is close to the Floquet eigenvalue of the
resonance state which can therefore mix easily with these bulk states.

\begin{figure}
\includegraphics[width=\columnwidth]{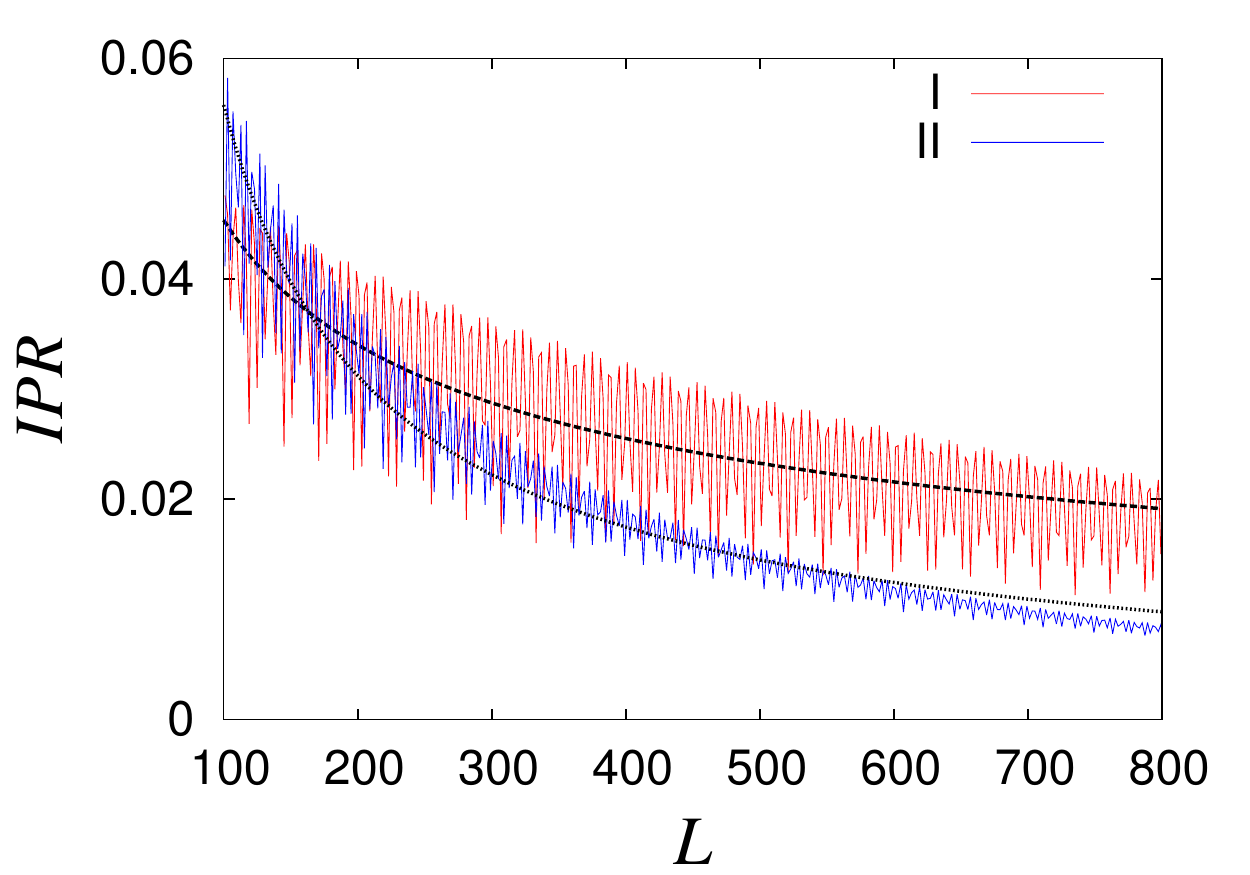}
\caption{The maximum IPR value of the eigenstates of the time evolution
operator as a function of the system size $L$ for (I) $\al=0.4$ and $T=2$ and 
(II) $\al=1.0$ and $T=2.5$. Curve I corresponding to the lower value of $\al$ 
has much larger fluctuations than curve II. The dotted lines show a fit of the 
form $IPR = a/L^b$ for the average IPR; for (I) the values $a=0.31$ and 
$b=0.41$ give the best fit; for (II) $a=2.65$ and $b=0.83$.} 
\label{fig:iprmax} \end{figure}

To see how the IPR of a bound or resonance state varies with the system
size, we study the maximum IPR versus $L$, taking $L$ to be odd, the kicking 
site to be at the middle, $L_c = (L+1)/2$, and open boundary conditions.
For $\al=0.4$ and $T=1$, we find that the maximum IPR is equal to $0.0936$ and
is independent of the system size in the range $101 \le L \le 799$. (We
have chosen this range so that $L$ is much larger than the decay length
$\lam$ of the central part of the state). This
size independence is a signature of a bound state. On the other hand,
for $\al=0.4$ and $T=2$, we find that the maximum IPR fluctuates significantly
for small changes in $L$ but on the average decreases as $L$ increases.
This is shown in Fig.~\ref{fig:iprmax}; the fluctuations demonstrate a 
sensitivity to the system size and confirm that it is a resonance rather 
than a bound state. We have chosen a fit of the form $IPR = a/L^b$;
we find that the best fit is given by the exponent $b= 0.41$ for $\al=0.4, ~T
=2$, and $0.83$ for $\al=1, ~T=2.5$.
This is to be compared with the IPRs of the bulk states which
decrease as $1/L$. Thus although the peak value of the wave function 
goes to zero as $L$ increases, the ratio of the peak value to the value 
of the wave function far from the peak grows with $L$. The value of the
exponent $b$ is not universal; we find that it depends on the values 
of $\al$ and $T$. However, it is smaller than 1 over a wide range of 
parameters and reasonably large system sizes,
implying that although the IPR of the resonance state decreases, 
the IPRs of the bulk states decrease even faster as $L$ increases. 

\begin{widetext}
\begin{center}
\begin{figure}
\begin{center}
\subfigure{\includegraphics[width=0.32\columnwidth]{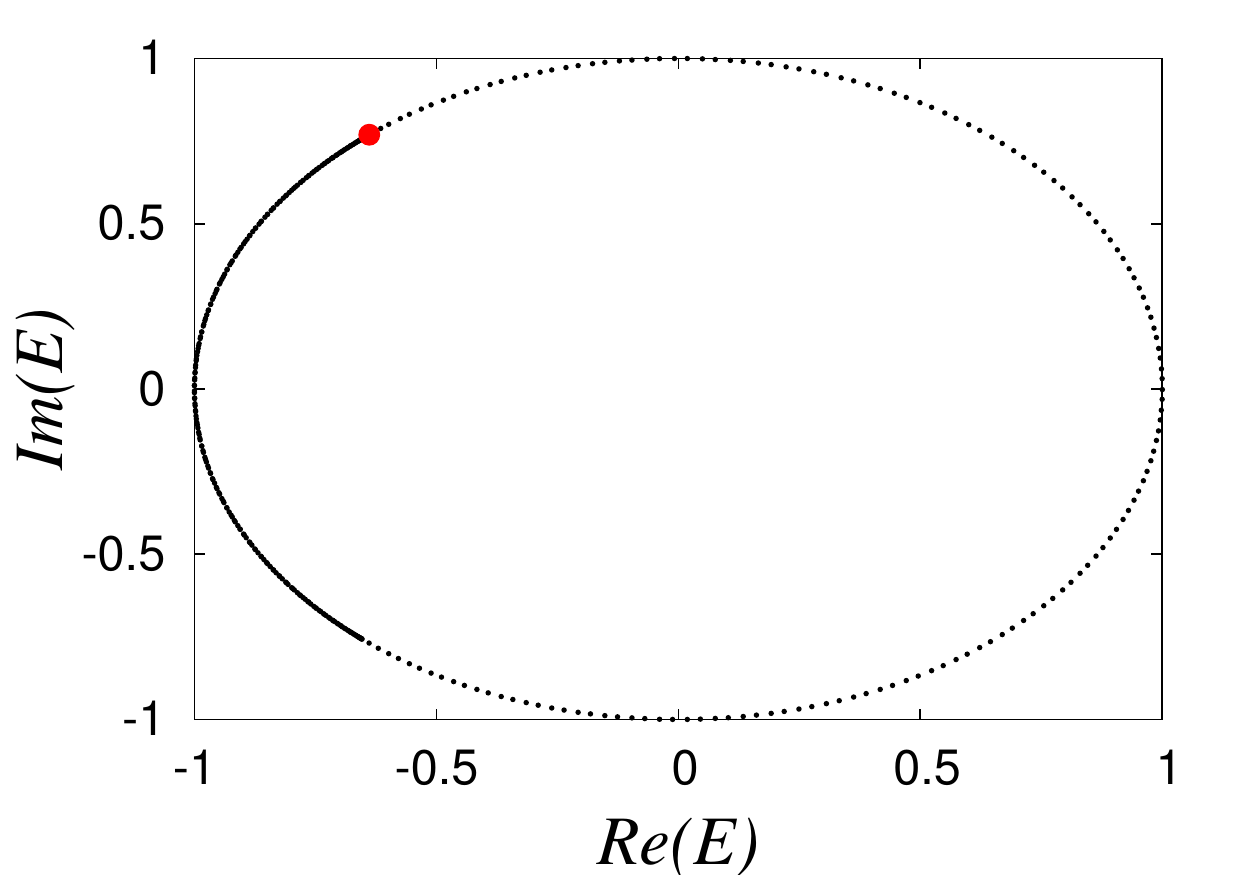}}
\subfigure{\includegraphics[width=0.32\columnwidth]{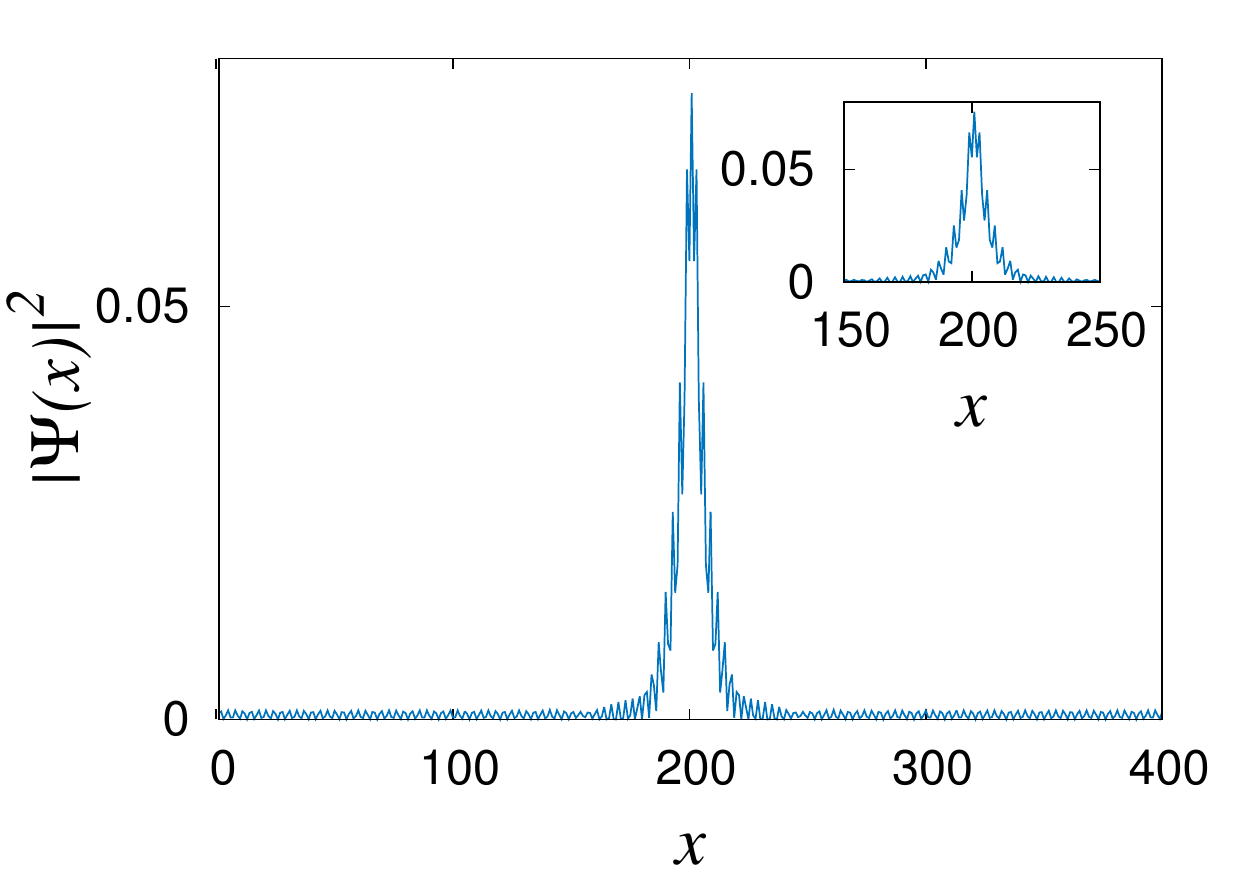}}
\subfigure{\includegraphics[width=0.32\columnwidth]{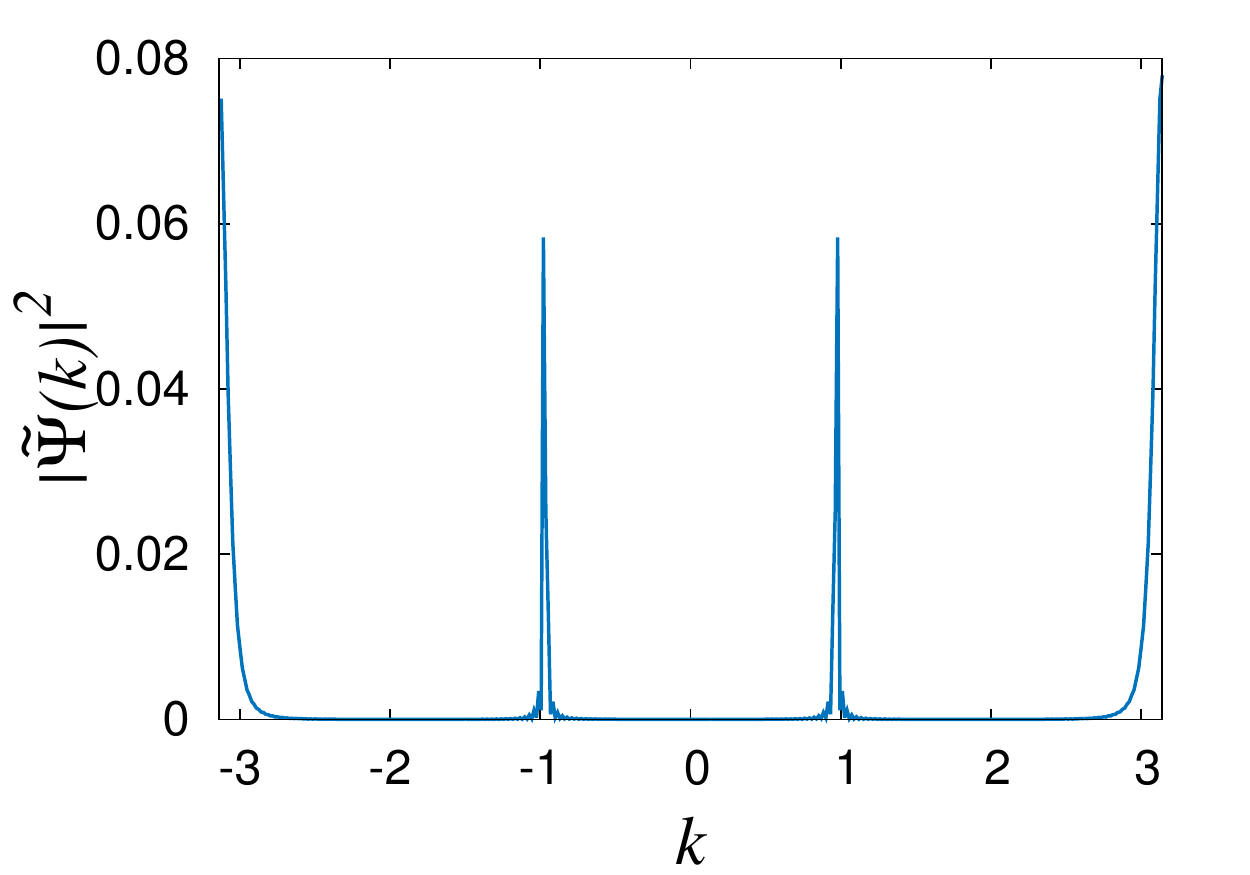}}
\end{center}
\caption{(Left) Eigenvalues of time evolution operator for a system with
401 sites in which $\de$-function kicks are applied to the $201$-th site
with strength $\al=0.4$ and time period $T=2$. (There are more eigenvalues
on the left side than on the right side; hence the plot looks solid on the
left and dotted on the right). There is a resonance state with
Floquet eigenvalue equal to $-0.6388 + 0.7694i$ shown by a large red dot. 
(Middle) Probability $|\psi(n)|^2$ of the resonance state. (Right) Square of 
the modulus of the Fourier transform, $|{\tilde \psi}(k)|^2$, of the resonance
state. It has peaks at both $k=\pm \pi$ and $k= \pm 0.967$, thus showing a 
substantial mixing with the bulk states.} \label{fig:resonance} \end{figure}
\end{center}
\end{widetext}

To summarize, we find that a bound state differs from a resonance in several 
ways.

(i) The Floquet eigenvalue of a bound state lies outside the 
continuum of the Floquet eigenvalues of the bulk states, while the Floquet 
eigenvalue of a resonance lies within the continuum of the bulk eigenvalues.

(ii) The wave function of a bound state is peaked at some point, 
decays rapidly away from that point, and essentially becomes zero beyond some 
distance. The wave function of a resonance is also peaked at some point and 
decays away from that point, but it does not become completely zero no matter 
how far we go; this is because it contains a non-zero superposition of some 
plane waves and therefore remains non-zero even far away from the peak.

(iii) If the system size is large enough, the properties of a
bound state, such as its IPR and the peak value of its wave function, 
become independent of the system size $L$ and the boundary conditions (for 
instance, whether we have periodic, anti-periodic or open boundary conditions). 
For a resonance, however, the IPR and peak value of the wave function depend 
sensitively on the boundary conditions and the value of $L$, and on the 
average they keep decreasing as $L$ is increased. This is because such a 
state contains some plane waves which sample the entire system, and the 
quasienergies of the plane waves is sensitively dependent on the boundary 
conditions and $L$. (We recall that if periodic boundary conditions are
imposed, the momentum of the plane wave states is quantized in units of 
$2\pi/L$. Hence the values of the momentum and therefore the quasienergies
$-2\ga \cos k$ depend on $L$).



Next, we introduce a time-independent on-site potential at the same site where 
the periodic kicking is being applied; this potential is given by
\beq H_V = V c^\dg_{L_c}c_{L_c}. \label{hv} \eeq
To investigate the effects of kicking, we again plot the maximum value of the 
IPR of all the eigenstates of the time evolution operator as a function of $T$
and $\al$. This is shown in the right panel of \Fig{fig:IPRphase} for $V=-4$.
We see a number of features in this plot including some straight lines; we now 
provide a qualitative understanding of these features. 
The bold lines within which the IPR is close to 1 is basically determined by 
whether the bound state mixes with the continuum states or not. Since
the bulk quasienergies lie between $-2\ga$ and $2\ga$, the bound state does 
not mix with the continuum states and therefore exists in the regions 
\bea \ep_b ~+~ \frac{\al}{T} &>& 2\ga, \label{cond1} \\
\ep_b ~+~ \frac{\al}{T} &<& -2\ga, \label{cond2} \eea
where $\ep_b$ is the bound state energy 
\beq \ep_b ~=~ -\sqrt{V^2+4\ga^2} \eeq
produced by an on-site potential $V < 0$. Similar to the condition 
in Eq.~\eqref{cond1} and using the fact that quasienergies are only defined 
modulo $2\pi/T$, we see that another line for the existence of a bound state 
is given by 
\beq \ep_b ~+~ \frac{\al}{T} ~>~ - ~\frac{2\pi}{T} ~+~ 2\ga. \label{cond3} \eeq
Between the two lines given in Eqs.~\eqref{cond1} and \eqref{cond2} and below 
the line given in Eq.~\eqref{cond3}, the bound state mixes with the bulk 
quasienergies and therefore turns into a resonance in the continuum. 

\section{Increase in conductance due to periodic kicking}
\label{sec_cond}

\begin{figure}
\includegraphics[width=\columnwidth]{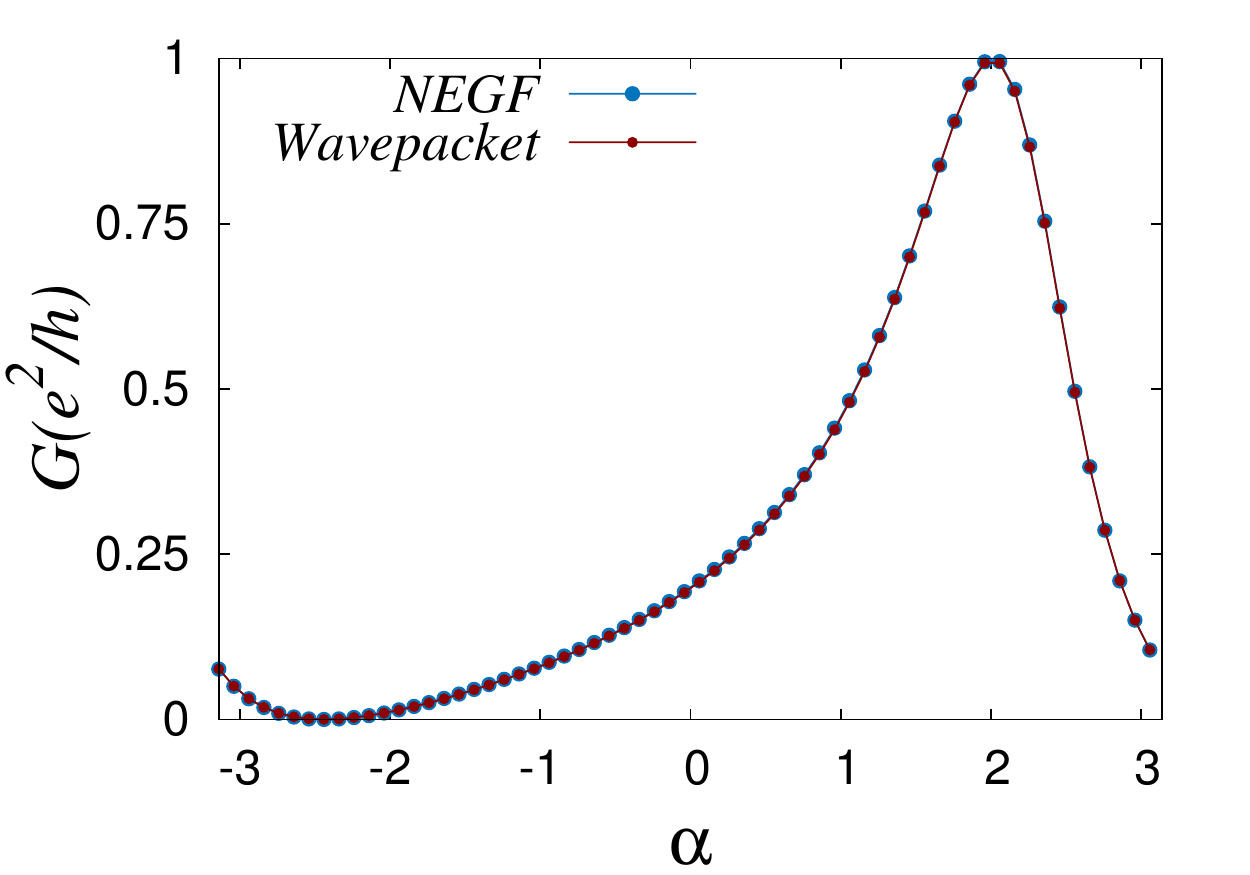}
\caption{Differential conductance $G = (e^2/h) {\cal T}$ vs $\alpha$ for 
a system with $V=-4$. We see that $\al$ can tune the conductance all the way 
from zero to 1. Here $L=401$, $k_c=\pi/2$, $L_o=50$, $\si=5$, and $T=0.5$.} 
\label{conduc0} \end{figure}

In the presence of only a time-independent on-site potential $V$, we have a 
bound state and a transmission probability $\cal T$ which is less than 1. We 
now ask if the transmission can be increased by periodic kicking at the same 
site where the potential $V$ is present. In Sec.~\ref{sec_ham} we saw that the 
transmission can get reduced when we introduce kicking. We now look at the 
opposite case where periodic driving can increase the transmission. 
In \Fig{conduc0}.
the differential conductance is shown as a function of $\al$ for a system
with $T=0.5$, $V=-4$ and $k_c = \pi/2$. The maximum transmission should occur 
at $\al=-VT$ which is equal to $2$ for the parameters used in \Fig{conduc0}. 
We see in that figure that this is indeed true and the system becomes
``transparent" when $\al=-VT$.

\section{Effects of Interactions}
\label{sec_inter}

\begin{figure}
\includegraphics[width=\columnwidth]{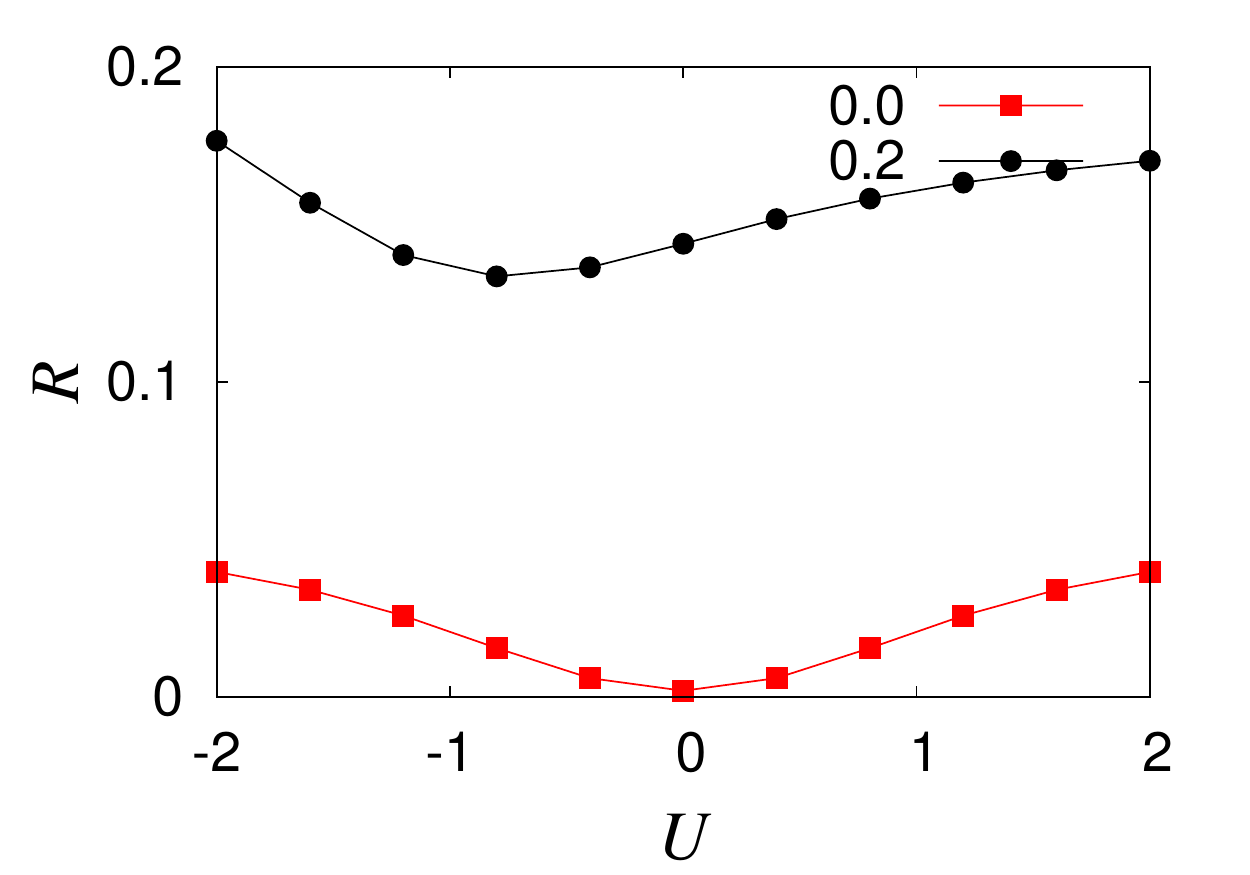}
\caption{Reflection probability ${\cal R}_\uparrow$ of a spin-up electron
for a wave packet in a 
system with an on-site interaction $U$ and kicking strengths $\al = 0$ (lower 
curve) and $0.2$ (upper curve). Even with $\al =0$ the wave packet is not 
completely transmitted in the presence of a finite $U$. This happens because 
there is a finite probability for the incoming wave packet to get 
trapped in a bound state which 
lives near the interacting site. The effect of $U$ acts asymmetrically in the 
presence of a finite $\al$. The effect of $U$ reduces with increasing $\al$ 
and at large values of $\al$, ${\cal R}_\uparrow$ is independent of $U$. We 
have taken $L=51$, $T=0.25$, $k_c =\pi/2$, $\si=4$, and $L_o=6$.} 
\label{fig:Ueff} \end{figure}

We now analyze the effects of interactions on the various aspects that we 
have discussed so far, namely, transport and the presence of bound states. 
We consider a system containing two species of 
electrons with up and down spins and a time-independent interacting term 
on the site $L_c$. The total Hamiltonian is
\bea H &=& - ~\ga ~\sum_{n=1}^{L-1} ~\sum_{\si = \uparrow,\downarrow} ~
(c^\dg_{n\si} c_{n+1,\si } + H.c.) \notag \\
&& + ~\al ~\sum_{m=-\infty}^\infty ~\de(t-mT) ~c^{\dg}_{L_c \si}
c_{L_c \si} \notag \\ 
&&+ ~U ~\hat{n}_{L_c \uparrow} \hat{n}_{L_c \downarrow}, \eea
where $\hat{n}_{L_c \si}$ is the number operator for electrons with spin $\si$ 
at site $L_c$. In order to investigate the effect of the interaction term, we 
begin with an initial wave packet which has two-particles in a singlet state of
$\psi_{\uparrow}(r)$ and $\psi_{\downarrow}(r)$; the form of the wave packet is
given in Eq.~\eqref{eqn:wavepa}.
We then study the effects of the interaction using exact diagonalization and 
wave packet evolution. The effect of $U$ is shown in \Fig{fig:Ueff}
where the reflection probability for an electron with spin-up, 
${\cal R}_\uparrow$, is shown as a function of $U$. [Given an amplitude $\psi 
(n_1, n_2)$ for a spin-up electron to be at $n_1$ and a spin-down electron to 
be at $n_2$, we define the reflection and transmission probabilities for a 
spin-up electron to be ${\cal R}_\uparrow = \sum_{n_1=1}^{L_c} \sum_{n_2=1}^L 
|\psi (n_1, n_2)|^2$ and ${\cal T}_\uparrow = \sum_{n_1=L_c+1}^L \sum_{n_2=1}^L
|\psi (n_1, n_2)|^2$, analogous to Eq.~\eqref{rt}. 
These satisfy ${\cal R}_\uparrow + {\cal T}_\uparrow = 
1$. We can similarly define ${\cal R}_\downarrow$ and ${\cal T}_\downarrow$; 
our choice of the form of the wave packet implies that ${\cal R}_\uparrow = 
{\cal R}_\downarrow$ and ${\cal T}_\uparrow = {\cal T}_\downarrow$]. Increasing
$\al$ makes $U$ less effective; this is because in the presence of kicking, 
the wave packet is small at the site $L_c$ and is therefore unable to sample 
the interaction. Note that in the presence of $\al$, the effects of $U$ and
$-U$ are different. This is because the driving produces an effective
on-site potential equal to $\al/T$; hence the total quasienergy of a state
with two electrons at site $L_c$ is $U + 2\al/T$. The minimum of this
energy (and hence the minimum of the reflection probability) occurs at 
a non-zero value of $U$.

As in the case of a non-interacting system, we find that Floquet bound states 
can also appear in the
presence of interactions. They have an interesting dependence on the time 
period $T$. If $U \gg \ga$, there will be a bound state in which both
electrons are at the site $L_c$, and the quasienergy of this state is 
$U + 2 \al/T$ in the presence of kicking. Since the energy of the 
bulk states of the two-electron system goes $-4\ga$ to $4\ga$, the bound state
will not mix with the bulk states if
\bea U ~+~ \frac{2\al}{T} &>& 4\ga, \label{cond4} \\
U ~+~ \frac{2\al}{T} &<& -4\ga. \label{cond5} \eea



\begin{figure}
\includegraphics[width=\columnwidth]{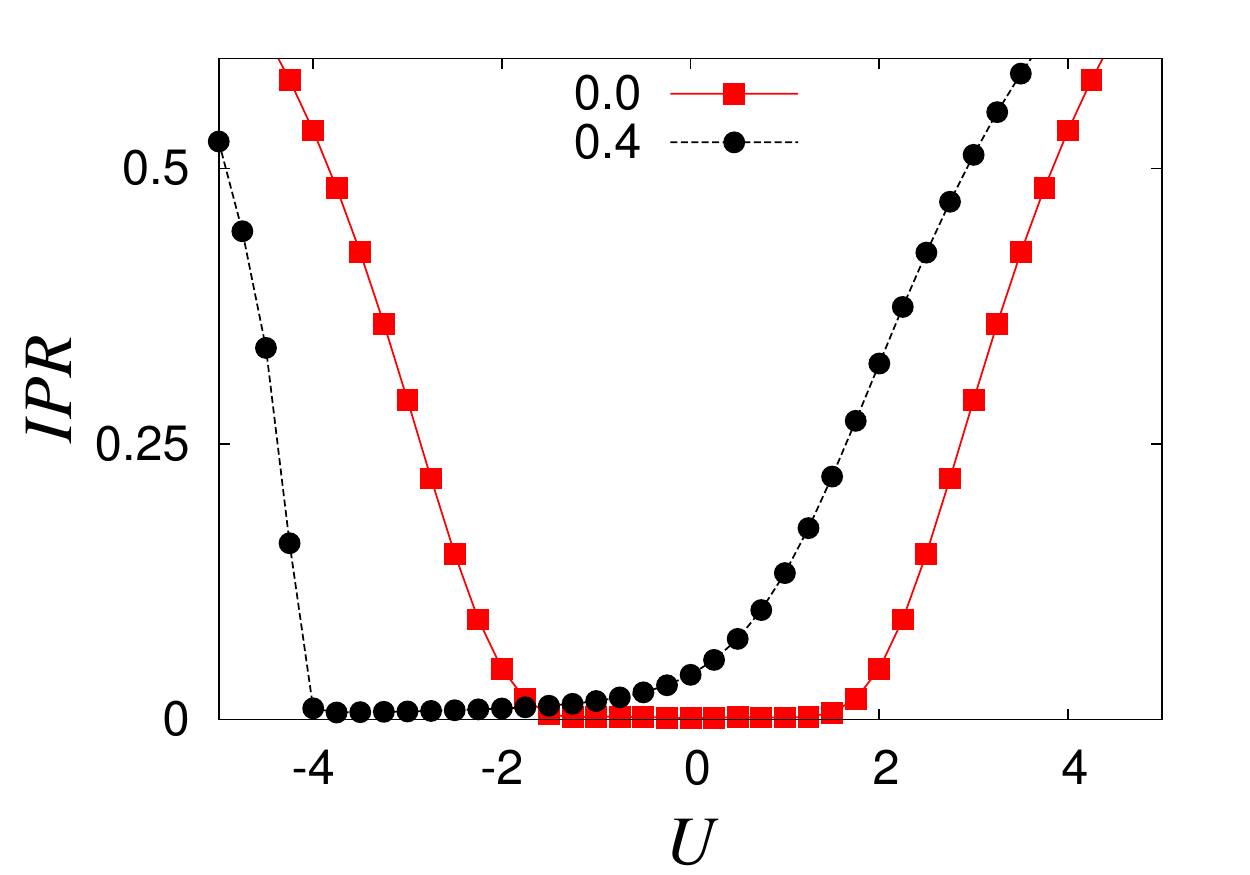}
\includegraphics[width=\columnwidth]{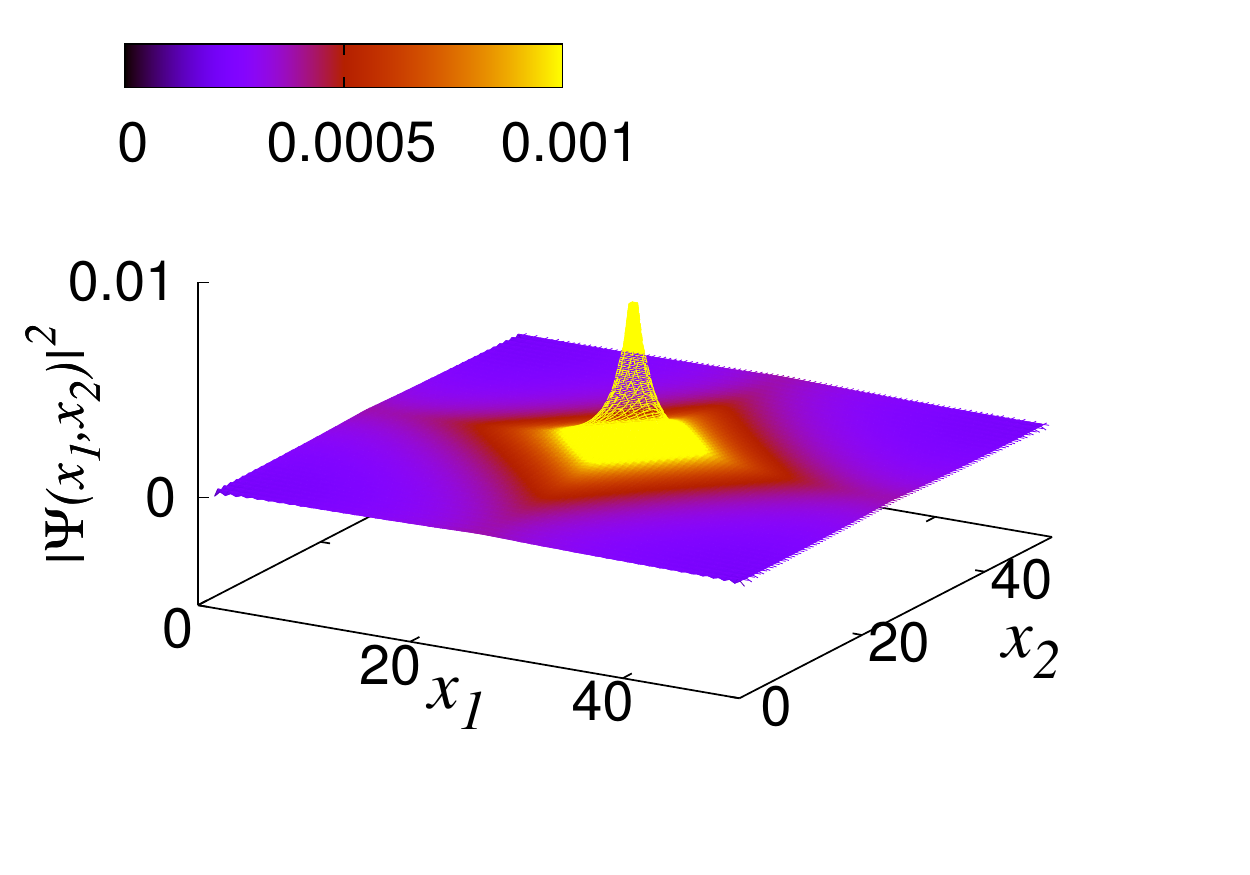}
\caption{(Top) The value of the two-body IPR for the state with the maximum 
IPR as a function of the Hubbard interaction $U$, with and without periodic
driving. We have taken $L=51$, $T=0.5$, $\al=0.4$ (black disks) and $\al=0$
(red squares). (Bottom) A two-particle bound state wave function for $L=51$, 
$T=0.5$, $\al=0.4$, and $U=1$ where a much more localized bound state is 
produced by the driving.} \label{IPRUchange} \end{figure}

We can calculate the IPR of a two-particle state and study its variation with 
$U$; this is shown in \Fig{IPRUchange}. In the absence of kicking ($\al =0$),
we find that for $|U| \lesssim 2$, the IPR is not very high suggesting a state 
which is not strongly localized, while for $|U| \gtrsim 2$, the IPR is large 
which suggests a strongly localized state. Interestingly, we find that a 
finite kicking strength (such as $\al = 0.4$) can turn a strongly localized 
state into a weakly localized one and vice versa, depending on the values of 
$U$ and $T$. A strongly localized two-particle bound state wave function is 
shown in \Fig{IPRUchange}.

\section{Conclusions}
\label{sec_concl}

In this paper we have studied the effects of periodic driving at one site in 
some tight-binding lattice models in one dimension. We have taken the driving
to be of the form of periodic $\de$-function kicks with strength $\al$ and 
time period $T$. We have studied how the kicking affects transmission across 
that site and whether it produces any bound states. 

The transmission (which is related to the differential conductance) has been 
calculated by constructing an incoming wave packet which is centered around 
a particular momentum and time evolving it numerically. The reflection and 
transmission probabilities are found by computing the total probabilities on 
the left and right sides of the kicking site after a sufficiently long time. 
The bound state is found by computing all the eigenstates of the time
evolution operator $U$ for one time period, and finding the eigenstate
with the maximum value of the inverse participation ratio; we look at the
corresponding wave function to confirm that it is indeed peaked near the
kicking site. 

When $T$ is much smaller than the inverse of the hopping $\ga$ and $|\al| \ll 
1$, the numerically obtained values of both the transmission probability and 
the Floquet quasienergy of the bound state are consistent with the fact that 
the kicking effectively acts like a time-independent 
potential equal to $\al/T$ at the special site. This is confirmed
by calculating the effective Hamiltonian and then using the non-equilibrium
Green's function method to compute the transmission; this is found to agree
well with the transmission found from wave packet dynamics if $T$ is small.
When $T$ is small and $\al = \pi$, we find that the transmission probability
is zero; this is because the effective hoppings between the kicked site and 
its two neighboring sites become zero, and it is related to the phenomenon of 
dynamical localization. On the other hand, when $T$ becomes comparable to 
$1/\ga$, the agreement between the transmissions found using wave packet 
dynamics and the effective Hamiltonian breaks down, showing that the effective
Hamiltonian no longer provides an accurate description of the system. 

We note that the effective breaking of a ``bond" in the Floquet 
description is a unique result and cannot be found in a static case. This 
feature is true even in higher dimensions and therefore provides a unique 
opportunity to experimentally simulate bond percolation problems \cite{Sen09} 
in cold atom or photonic systems. In such systems local $\delta$-function 
kicking can be implemented at randomly chosen sites in the system and their 
effect on the localization physics can be investigated.

A bound state can appear only if its quasienergy does not lie the continuum 
of the quasienergies of the bulk states going from $-2\ga$ to $2\ga$ modulo 
$2\pi/T$. If the Floquet quasienergy of the would-be bound state lies within 
the continuum of the quasienergies of the bulk states (this necessarily 
happens if $T > \pi/(2\ga)$ but it can also happen for certain values of 
$\al$ if $T < \pi/(2\ga)$), the bound state ceases to exist. However, we then 
find that in certain ranges of values of $\al$ and $T$, there is a state which 
can be described as a resonance in the continuum. The wave function of such a 
state consists of a superposition of a strongly peaked part which resembles
a bound state and a plane wave part which does not decay even far away from 
the kicking site. Further, the IPR of this state is sensitively dependent on 
the system size and boundary conditions and it gradually decays as the system 
size is increased. This behavior is in contrast to a bound state whose IPR 
becomes independent of the system size and boundary conditions when the 
system size is larger than the decay length.

Next, we have studied what happens if there is a time-independent potential 
$V$ at a single site and periodic $\de$-function kicks are applied to the 
same site.
Separately both $V$ and the kicks reduce the transmission from unity and can 
produce bound states. When both of them are present, we get a complex pattern 
of regions in the $\al-T$ plane where bound states are present. These regions 
can be understood using a simple condition that the sum of the effective 
on-site potential $\al/T$ due to the kicking and the energy of the bound 
state produced by $V$ alone should not lie within the continuum of the 
quasienergies of the bulk states. Further, if $V$ and $\al/T$ have opposite
signs, their effects can partially cancel each other and the transmission
probability can be higher than if only one of them was present.

Finally, we have studied a model with spin-1/2 electrons where there is a 
Hubbard interaction of strength $U$ at a single site and periodic kicks are 
applied to the same site. We numerically study wave packet dynamics starting 
with an initial wave packet which contains two electrons in a spin singlet
state. In the absence of kicking, a state in which both particles are
at the special site has an energy $U$. This has a similar effect as an 
on-site potential for the model of spinless electrons; the transmission
probability is therefore reduced from 1 for any non-zero value of $U$,
and it has the same value for $U$ and $-U$. When we introduce kicking, the 
effective potential for two particles at the special site is given by the 
sum of $U$ and $2\al/T$. Hence the transmission probability will be higher 
when $U$ and $2\al/T$ have opposite signs, and will therefore not be symmetric 
under $U \to - U$. We also find that a bound state can appear if its
quasienergy does not lie within the continuum of bulk quasienergies.
When $U$ is non-zero, we find that kicking can convert strongly
localized states to weakly localized ones and vice versa.

We end by pointing out some directions for future studies.

 (i) In this paper we have only examined systems with one or two 
particles. It may be interesting to study a thermodynamic system with a finite
filling fraction of particles. One can then investigate if, for example,
the model of spin-1/2 electrons with both an interaction $U$ and periodic
kicking at the same site can show a Kondo-like 
resonance~\cite{hewson97,kaminski00}.
Related problems have been studied in Refs.~\onlinecite{heyl10,iwahori16}.

(ii) It may be interesting to look at the effects of heating. It is 
known that a system generally heats up to infinite temperature when there are
interactions and periodic driving at all the
sites~\cite{bilitewski15,genske15,kuwahara16}. However, if interactions and
periodic driving are both present in only a small region as considered in this
paper, it is not known if the system will heat up indefinitely at long times.

(iii) The effects of periodic kicking at more than one site,
possibly with different strengths and phases, would be interesting to study. 
It is known that harmonic driving at two sites with a phase difference
can pump charge (see Ref.~\onlinecite{soori10} for references). We 
therefore expect that the application of $\de$-function kicks at two sites 
may also pump charge. In addition, we can study what kinds of bound states are
generated in such a system.

There has been an increasing interest in understanding the dynamics 
of a single impurity or of electrons in a quantum dot under the periodic 
modulation of some parameter. This is motivated both by theoretical 
considerations such as the effect of such a modulation on the Kondo 
effect \cite{iwahori16, Suzuki17} and by advances in cold atom 
experiments which allow for the imaging and modulation of systems up to 
single site resolution \cite{Fukuhara13, Fukuhara132, Hild14}. The results 
presented in this manuscript describe many interesting phenomena which are 
realizable due to an interplay of impurity physics and dynamical modulation 
of some parameter in the Hamiltonian. It will be interesting if such effects 
can indeed be observed in cold atom or mesoscopic systems.
\vspace*{.8cm}

\section*{Acknowledgments}

A.A. thanks Council of Scientific and Industrial Research, India for funding 
through a SRF fellowship. D.S. thanks Department of Science and Technology, 
India for Project No. SR/S2/JCB-44/2010 for financial support.

\end{document}